\documentclass[iop]{emulateapj}
\usepackage{ulem}

\shorttitle{SATELLITE FORMATION AROUND GIANT PLANETS}

\shortauthors{Ogihara et al.}

\begin{document}

\title{\textit{N}-BODY SIMULATIONS OF SATELLITE FORMATION AROUND GIANT PLANETS: ORIGIN OF ORBITAL CONFIGURATION OF THE GALILEAN MOONS}

\author{Masahiro Ogihara}
\affil{Nagoya University,
Furo-cho, Chikusa-ku, Nagoya, Aichi 464-8602, Japan}
\email{ogihara@nagoya-u.jp}
\and
\author{Shigeru Ida}
\affil{Tokyo Institute of Technology,
Ookayama, Meguro-ku, Tokyo 152-8551, Japan}
\email{ida@geo.titech.ac.jp}

\begin{abstract}
As the number of discovered extrasolar planets has been increasing,
diversity of planetary systems requires studies of new formation scenarios.
It is important to study satellite formation in circumplanetary disks,
which is often viewed as analogous to formation of rocky planets in protoplanetary
disks.
We investigated satellite formation from satellitesimals around giant planets
through \textit{N}-body simulations that include gravitational interactions with a 
circumplanetary gas disk.
Our main aim is to reproduce the observable properties of the Galilean satellites
around Jupiter through numerical simulations, as previous 
\textit{N}-body simulations have not explained the origin of the resonant
configuration.
We performed accretion simulations based on the work of \citet{sasaki_etal10},
in which an inner cavity is added to the model of \citet{canup02,canup06}.
We found that several satellites are formed and captured in mutual mean motion 
resonances outside the disk inner edge and are stable after rapid disk gas
dissipation, which explains the characteristics of the Galilean satellites.
In addition, owing to the existence of the disk edge, a radial compositional
gradient of the Galilean satellites can also be reproduced.
An additional objective of this study is to discuss orbital properties of formed satellites
for a wide range of conditions by considering large uncertainties in model parameters.
Through numerical experiments and semianalytical arguments, we determined that 
if the inner edge of a disk is introduced,
a Galilean-like configuration in which several satellites are captured into a 2:1 resonance
outside the disk inner cavity is almost universal. In fact, such a configuration
is produced even for a massive disk $\gtrsim 10^4~{\rm g~cm^{-2}}$ and 
rapid type I migration.
This result implies the inevitability of a Galilean satellite formation in addition to
providing theoretical predictions for extrasolar satellites. That is, we can predict a
substantial number of exomoon systems in the 2:1 mean motion resonance
close to their host planets awaiting discovery.

\end{abstract}

\section{INTRODUCTION}
\label{sec:intro}
Origins of satellites around outer giant planets such as Jupiter and 
Saturn are important for determining the history of these objects, 
and their existence holds clues to the origins of such planets.
Satellites around Jupiter and Saturn, in fact, are suitable 
targets for formation theories because their orbital parameters and 
compositions are well documented (e.g., \citealt{seidelmann92})
compared to those of exoplanets.
Therefore, we have developed a formation theory that compares
physical parameters.

Besides satellites in the solar system, 
those in extrasolar systems, known as exomoons, are an additional focus 
of this study.
Exomoons have recently attracted great interest
because of their habitability.
The number of detected extrasolar giant planets is 
much larger than rocky exoplanets, with the exception of Kepler 
candidates, and some orbit within the habitable zone
 (HZ),  i.e., an orbital region in which the stellar flux is sufficient 
to maintain liquid water on the surface of a planet 
(e.g., \citealt{kasting_etal93}).
If rocky satellites orbit giant planets in the HZ, they 
are potentially habitable. 
According to studies on the possibility of 
life-bearing moons (\citealt{williams_etal97}, \citealt{kaltenegger00}),
satellites around extrasolar giant planets in the HZ might be 
habitable if their size is sufficient ($\gtrsim 0.2~M_{\oplus}$).
An obvious exception is Titan, which has a dense atmosphere even
though its mass is $\simeq 0.02~M_{\oplus}$.

A detailed study on the detectability of habitable exomoons 
by \citet{kipping_etal09} determined that exomoons in the HZ 
down to $0.2~M_{\oplus}$ may be detected by transit
timing effects, including transit timing variations and transit duration 
variations, with the expected performance of the 
\textit{Kepler space telescope}. In fact, an ongoing observational project uses
photometry data of \textit{Kepler} to detect and investigate exomoon signals 
\citep{kipping_etal12}. 
In addition, the possibility of screening the atmosphere of 
exomoons for habitability with transmission spectroscopy
has been discussed by \citet{kaltenegger10}, who discovered
that the number of transits needed to detect 
biomarkers on an Earth-like exomoon under idealized 
conditions and viewing geometry is feasible using the 
\textit{James Webb Space Telescope} (\textit{JWST})
for the sample of the closest M stars.
Thus, it is important to predict the characteristics of satellites,
such as mass, composition, and orbital location, 
that are likely to exist around extrasolar giant planets. 

Regular satellites, which follow prograde and relatively close
orbits up to tens of planetary radius with little orbital eccentricities
and inclinations, are believed to have formed 
as by-products of planets' formation 
within a circumplanetary accretion disk (e.g., \citealt{lunine82}).
Circumplanetary disk models currently fall into two categories:
(i) solids-enhanced minimum mass (SEMM) model
(\citealt{mosqueira03a,mosqueira03b}; 
\citealt{estrada_etal09}), in which a stationary disk is assumed; 
and (ii) gas-starved disk model
(\citealt{canup02,canup06,ward_canup10}), also known as a slow inflow disk,
in which an actively supplied gaseous disk is considered.

In the SEMM model, a disk is composed of two different zones:
an optically thick inner region with a peak surface density 
$\sim 10^5 \rm{~g~cm}^{-2}$ and an optically thin outer region 
extended to a fraction of the Hill radius of the planet 
($\sim 0.2~r_{\rm H}$).
The transition from the inner disk to the outer disk is assumed 
to occur at the location between Ganymede and Callisto 
for the Jovian disk. 
In this model, the disk is static, that is there is no inflow
from circumstellar orbits. The inner high density region with very low
turbulent viscosity ($\alpha \simeq 10^{-6}-10^{-5}$) is practically inviscid.
Solid components of the disk are supplied by ablation and capture 
of planetesimal fragments passing through the disk.
In the gas-starved disk model, however, a low-mass
viscously evolving disk with a peak surface density of $\sim 100
\rm{~g~cm}^{-2}$ is considered. This disk is continuously supplied
by the ongoing inflow of gas and dust particles from the protoplanetary disk.
Disk conditions evolve in a quasi-steady state mode in response 
to the rate of gas accretion and turbulent diffusion,
leading to lower disk gas densities than in the SEMM model.
In this study, we 
adopt a fiducial model for gas surface density
based on the gas starved disk model ($\sim 100\rm{~g~cm}^{-2}$).
Note that if low disk viscosity or rapid gas inflow is assumed, 
the gas surface density can become higher than $100\rm{~g~cm}^{-2}$
(see Equations~(\ref{eq:surface_density}) and (\ref{eq:fg})).

\begin{deluxetable}{lccccc}
\tablecolumns{5}
\tablewidth{0pc}
\tablecaption{Properties of the Galilean Satellites}
\startdata
\hline
\hline
Satellites & $ a~(R_{\rm P})$ & $M~(10^{-5}~M_{\rm P})$ &$\rho~({\rm g~cm^{-3}})$& $e$ &$i$~(rad)\\
\hline
Io & 5.9 & 4.7 & 3.53&0.0041&0.0007\\
Europa & 9.4 & 2.5 & 2.99&0.01&0.0082\\
Ganymede & 15.0 & 7.8 & 1.94 &0.0015&0.0034\\
Callisto & 26.4 & 5.7 & 1.83&0.007&0.0049
\enddata
\tablecomments{Values are taken from \citet{schubert_etal04} and \citet{yoder95}.
$a$, $M$, $\rho$, $e$, and $i$ are the semimajor axis, mass, material density, 
eccentricity, and inclination of the satellites, respectively.}
\label{tbl:galilean_satellites}
\end{deluxetable}

Our main focus in this paper is the formation of Galilean-like satellites,
although we extend our theory to exomoons and also comment on the Saturnian
system later in this paper. Physical properties of the Galilean satellites are shown
in Table~\ref{tbl:galilean_satellites}.
Orbital and compositional characteristics can 
provide valuable constraints on formation scenarios.
The orbital resonances among the satellites 
Io, Europa, and Ganymede present a fascinating
dynamic system. Io-Europa and Europa-Ganymede 
are in the 2:1 mean motion resonance, which causes
their successive conjunctions to occur near the same 
longitude. These satellites are also in the Laplace resonance, which denotes
the 1:1 commensurability between the rates of motion 
of the Io-Europa and Europa-Ganymede conjunctions;
thus, a triple conjunction never occurs.
These resonances suggest that the Galilean satellites
underwent orbital migration either during or after 
their formation.
With regard to composition, a progressive increase 
in satellite ice mass fraction with increasing distance 
from Jupiter is observed. Io is composed of rock, and 
Europa, Ganymede, and Callisto are believed to contain approximately
8\%, 45\%, and 56\% ice and water by mass, respectively
(e.g., \citealt{schubert_etal04}). 
In addition, Callisto's interior structure allows it to constrain
its accretion timescale. 
Data from the \textit{Galileo spacecraft} suggest that Callisto appears
to be largely undifferentiated,
provided that the satellite is in hydrostatic equilibrium
(e.g., \citealt{anderson_etal01}).
Therefore, Callisto must have avoided melting in its entire history.
Estimates of the temperature increase associated with accretional heating 
show that 
Callisto could have remained unmelted during formation if its accretion
was on a timescale longer than $5 \times 10^5$~years \citep{barr08}.
In addition, eccentricities and inclinations are considerably small,
which are considered to be damped by the surrounding gas nebula
and/or tidal dissipation within the satellites.

Formation of the Galilean satellites has been investigated 
from several perspectives.
\citet{canup06} performed \textit{N}-body simulations of 
satellite accretion from satellitesimals in a starved disk that
included the effects of a gas disk.
Because the type I decay and accretion timescales 
are shorter than the disk lifetime, several generations of
satellite formation and migration are repeated before gas 
disk dissipation.
As the infall of gas and solid materials wanes because of global 
depletion of the circumstellar disk, the circumplanetary disk
is depleted through viscous spreading and/or photoevaporation.
Thus, the current satellites are the survivors of the final 
generation.
\citet{canup06} also found that the total satellite mass fraction to the host 
planet is regulated 
to $\sim 10^{-4}~M_{\rm P}$ ($M_{\rm P}$; mass of the planet), 
which is only weakly dependent on model
parameters and is consistent with Jovian and Saturnian systems.

Several authors attempt to clarify the origin of resonant states. 
 \citet{yoder79} and \citet{yoder81} proposed that Io captured
Europa into the 2:1 resonance through differential orbital expansion 
due to gravitational tides raised on Jupiter. With continuing orbital 
expansion, Europa eventually encountered the 2:1 mean motion
commensurability with Ganymede, resulting in capture into 
the Laplace resonance.
In contrast, \citet{greenberg87} and \citet{peale02} advocated
a primordial origin of the resonant states through
differential inward type I migration in the gaseous disk.
Since the type I migration timescale is inversely proportional to
the satellite mass, the largest satellite, Ganymede, underwent
the most rapid drift and caught up with the inner satellites, 
leading to capture into the resonances.
However, \citet{canup06} performed detailed \textit{N}-body simulations, 
which is the first \textit{N}-body study of satellite accretion, and it is 
suggested that resonant states cannot be produced in their 
calculations\footnote{\citet{canup06} did not analyze whether the
satellites evolved in resonances or not, but we performed \textit{N}-body
simulations under the same condition as \citet{canup06} and confirmed
that resonant relationships were hardly established.}.
It seems that
the initial orbital configuration of \citet{peale02} cannot be established during
Canup \& Ward's \textit{N}-body simulations, because when a Ganymede-mass
satellite forms in outer region, satellites that reside in inner region do not grow
to the mass comparable to those of Io and Europa.

As shown in recent studies on formation of close-in super-Earths
\citep{terquem07,ogihara09,papaloizou10},
migrating bodies can be trapped in mean motion resonances 
near the disk inner edge.
\citet{sasaki_etal10} introduced the concept of the disk inner cavity in their
investigation of satellite formation and found that four or five satellites
are generally formed by being trapped in resonances, which agrees well
with the Galilean satellites. 
\citet{sasaki_etal10} adopted a semianalytical method rather than 
direct orbital integration and performed Monte Carlo 
simulations. Although this approach provides a powerful means for 
overall understanding of satellite formation,
it is not adequate for effectively discussing detailed features of resonant trapping, 
because particularly in the formation of Galilean satellites, similar-sized satellites 
interact with each another, and multiple bodies are eventually trapped
in resonances. 
Several studies outlined the prediction of capture 
probability for resonances as a function of initial particle eccentricity 
in the adiabatic limit, where the migration timescale is much
longer than the resonant libration timescale 
(e.g., \citealt{murray99}). Recently, \citet{quillen06} and \citet{mustill10}
studied a rapid migration case in the 
restricted three-body problem in the limit 
of low initial particle eccentricity
and semianalytically obtained capture
probability using a Hamiltonian model. However, it is uncertain whether
their formulae can be used in a case in which bodies have
comparable masses and the effect of eccentricity damping is 
included.
Therefore, the \textit{N}-body simulation is the only currently 
available method to accurately examine the detailed features of resonant 
trapping in systems in which
multiple bodies with comparable masses interact with each other.

Thus, we conduct \textit{N}-body simulations of satellite
accretion in a gas-starved disk \citep{canup02} by adding an inner cavity
to the original model, to clarify the formation scenario with the disk edge.
In particular, we assess the probability of primordial formation of the 
resonance relationship. Moreover, we attempt to reproduce additional properties of 
Galilean satellites such as the total number and the compositional gradient.

Furthermore, we examine the dependence of model parameters on the 
properties of formed satellites. Recent studies on disk-planet interactions
have suggested that torque exerted on a planet from a disk can
be significantly altered because of disk structure \citep{masset_etal06} and radiative effects
(e.g., \citealt{paardekooper_etal10}), resulting in a discrepancy 
with the type I migration speed predicted by linear calculations
\citep{tanaka_etal02}.
In an actively supplied disk, in which
equilibrium between gas inflow and viscous diffusion is assumed, 
gas surface density depends on the inflow flux and gas viscosity.
Given such uncertainties, we discuss satellite formation under
a wide range of parameters. The results of this study are instrumental for effective 
prediction of orbital properties of extrasolar satellite systems.

In Section~\ref{sec:model} of this paper, we detail our model and numerical method; 
in Section~\ref{sec:n-body}, we present the results of \textit{N}-body simulations;
in Section~\ref{sec:parameters}, we discuss the dependence of model parameters
on the properties of formed satellites;
in Section~\ref{sec:saturn}, we briefly discuss satellite formation for the Saturnian
system; and in Section~\ref{sec:conclusion}, we summarize our conclusions and the implications
of this study.

\section{MODEL AND CALCULATION METHOD}
\label{sec:model}
\citet{sasaki_etal10} considered two different models that correspond to
Jovian and Saturnian systems: Surviving satellites around Jupiter were
formed in a relatively massive protosatellite disk with an inner disk cavity, 
and those around Saturn were formed in a less massive
and cooler disk without an inner disk cavity.
These disk conditions are inferred from gap openings along the host planet's
orbit in the circumstellar protoplanetary disk and through observations of 
classical T-Tauri stars (CTTSs) and weak-line T-Tauri stars (WTTSs).

Gap openings require that both viscous and thermal conditions be satisfied 
\citep{lin85}. The critical masses 
are expressed as \citep{ida04}:
\begin{eqnarray}
M_{g,\rm{vis}} \simeq 30 \left(\frac{\alpha}{10^{-3}}\right)
\left(\frac{a}{1~\rm{AU}}\right)^{1/2} \left(\frac{M_*}{M_\odot}\right)
~M_\oplus, \\
M_{g,\rm{th}} \simeq 120\left(\frac{a}{1~\rm{AU}}\right)^{3/4}
\left(\frac{M_*}{M_\odot}\right)~M_\oplus,
\end{eqnarray}
where $M_{g,\rm{vis}}$ and $M_{g,\rm{th}}$ are the viscous and the thermal
conditions, respectively. Although these masses have some uncertainties,
$M_{g,\rm{vis}}$ and $M_{g,\rm{th}}$ are larger at Saturn's orbits than at
Jupiter's orbit. Since Saturn's mass is three times less than Jupiter's one,
it is plausible that Jupiter opened a gap to halt its growth 
while Saturn did not and Saturn's growth was terminated by global dissipation 
of the solar nebula \citep{sasaki_etal10}.
As will be discussed in Section~\ref{sec:disk_model},
gap openings lead to rapid depletion of protosatellite disks; 
satellites formed in a relatively massive protosatellite disk could survive around Jupiter,
while those formed in a late-stage less massive disk could survive around Saturn.

\citet{herbst_mundt05} observed that
spin periods of young stars show bimodal distribution in peaks at 
about 1 week and 1 day. 
They suggested that the stellar magnetic field
of the 1-week-period stars is coupled with the circumstellar disk and is sufficiently strong
to transfer spin angular momentum to the disk and open a disk inner cavity,
whereas the disks around the 1-day-period stars do not have the cavity
(see also \citealt{hartmann02}).
They also suggested that the 1-week and 1-day stars may correspond
to CTTSs and WTTSs, respectively, enabling 
massive CTTS disks with a cavity to evolve to
less massive WTTS disks without a cavity. In this evolution analogy,
protosatellite disks may have a cavity in relatively early stages that 
disappears as the disk evolves, although this scenario includes a large uncertainty.
Because the current spin rate of Jupiter is much slower than its break-up spin
rate, magnetic coupling between Jupiter and the
circum-Jovian disk is an obvious assumption. In fact, \citet{takata_stevenson96} showed that
the circum-Jovian disk may have contained an inner cavity.
The satellites formed in such a disk survived in case of Jupiter.
The circum-Saturnian disk may once have contained a similar disk in early stages.
However, if the disk evolved to a less massive disk without a cavity,
the satellites formed in the massive disk would have fallen onto Saturn, and
the surviving satellites would have formed in the late-stage disk.
We comment on the formation of the Saturnian system in Section~\ref{sec:saturn}.
In the rest of this paper, we focus on the model for Jupiter.

\subsection{Disk Model}
\label{sec:disk_model}
\subsubsection{Gas Surface Density and Disk Temperature}
\label{sec:gas_density}
As stated in Section~\ref{sec:intro}, we adopted the actively supplied 
gas-starved disk model in our
calculations.
Our disk model is based on that by \citet{sasaki_etal10}, in which 
the disk inner cavity is added to the model by \citet{canup02} with 
other slight modifications. In this section, we briefly summarize the 
model by \citet{sasaki_etal10}.

For the case in which gas inflow from the circumstellar
disk is limited to the region between
$r_{\rm in}$ and $r_{\rm c}$,
\citet{canup09} suggested that the inner and outer radial 
boundaries of the inflow region for the proto-Jovian disk are 
$4~R_{\rm J}$ and $34~R_{\rm J}$, respectively. This theory is
consistent with the three-dimensional hydrodynamic
simulation by \citet{machida09}, i.e., $r_{\rm c} \sim 22~R_{\rm J}$,
where $R_{\rm J}$ is the physical radius of Jupiter.
As adopted by \citet{canup06} and \citet{sasaki_etal10}, we used $r_{\rm c} 
= 30~R_{\rm P}$ in this study.
The total inflow rate is
expressed as $F_{\rm P} = M_{\rm P}/\tau_{\rm G}$,
where $\tau_{\rm G}$ is the gas inflow timescale;
then the infall flux per unit area is $F_{\rm in} = 
F_{\rm P}/\pi(r_{\rm c}^2 - r_{\rm in}^2) \simeq
F_{\rm P}/\pi r_{\rm c}^2$.
Here we ignore the radial dependence of 
the inflow because the resolution of the current hydrodynamic simulations 
is insufficient to constrain the radial
dependence of the infall, as will be discussed in Section~\ref{sec:saturn}.

A disk is formed with the inflowing gas, which diffuses
viscously. If the viscous diffusion timescale is shorter 
than the characteristic timescale over which the inflow
changes, the gas disk can be described as a steady
accretion disk, in which the inflow and viscous diffusion 
are equilibrated. According to the steady-state disk model
derived by \citet{canup02}, the gas surface density of the 
disk is approximately given by (Appendix~\ref{app:disk})
\begin{eqnarray}
\label{eq:surface_density}
\Sigma_{\rm g} &\simeq& 0.55 \frac{F_{\rm P}}{3\pi\nu}\nonumber\\
&\simeq& 100 f_{\rm g} \left(\frac{M_{\rm P}}{M_{\rm J}}\right)
\left(\frac{r}{20R_{\rm P}}\right)^{-3/4}
\left(\frac{R_{\rm P}}{R_{\rm J}}\right)^{-3/4} {\rm ~g~cm^{-2}},
\end{eqnarray}
\begin{eqnarray}
\label{eq:fg}
f_{\rm g}  \equiv \left( \frac{\alpha}{5\times10^{-3}}\right)^{-1}
\left(\frac{\tau_{\rm G}}{5\times10^6 {\rm ~yr}}\right)^{-3/4},
\end{eqnarray}
where $\nu$, $M$, $r$, and $R$ are the viscosity, mass, 
radial distance from the planet, and physical radius of the
bodies, respectively.
The subscripts ``P'' and ``J''
denote quantities of the host planet and Jupiter, respectively.
Because the magnitudes of turbulent viscosity and 
the gas inflow rate are not well determined, we introduce
the scaling factor $f_{\rm g}$ for gas surface density. 
We adopt an alpha model for the disk viscosity \citep{shakura73}
$\nu = \alpha c_s H \simeq \alpha c_s^2 /\Omega_{\rm K}$,
where $H$ and $\Omega_{\rm K}$ are the disk scale height and
the Keplerian angular velocity, respectively, and
the sound velocity $c_s$ is derived thorough the temperature 
distribution $T$. $T$ is determined by the balance between viscous 
heating and blackbody radiation (Appendix~\ref{app:disk}) as follows:
\begin{eqnarray}
\label{eq:temp}
T &\simeq& 160 \left(\frac{M_{\rm P}}{M_{\rm J}}\right)^{1/2}
\left(\frac{\tau_{\rm G}}{5 \times 10^6 {\rm ~yr}}\right)^{-1/4}
\nonumber\\ && \times \left(\frac{r}{20 R_{\rm P}}\right)^{-3/4}
\left(\frac{R_{\rm P}}{R_{\rm J}}\right)^{-3/4} {\rm K}.
\end{eqnarray}

\subsubsection{Gas Dissipation and Disk Inner Edge}
Gas surface density depends on the inflow rate;
thus, the decay of the inflow leads to a decrease in  
gas surface density (and disk temperature).
In our calculations, we generally assume an exponential decay 
of the inflow rate as $F_{\rm P} \times \exp(-t/\tau_{\rm dep})$,
where the decay timescale $\tau_{\rm dep}$ is  set to 
be $10^6$ years. We therefore adopt an exponential decay of 
gas surface density with a dissipation timescale of $10^6$ years.
Altough more accurate dependence of the inflow rate on 
the surface density is represented by $\Sigma_{\rm g} \propto F_{\rm P}
/\nu \propto F_{\rm P}^{3/4}$,
we used the same decay timescale for simplicity.

Although infall flux decay associated with global dissipation
of a protoplanetary disk is $\sim 10^6$ years, we introduce a significantly more
rapid decay associated with inflow truncation by the gap opening in the
protoplanetary disk.
After the gap opening, the protosatellite disk would be
rapidly depleted on the viscous diffusion timescale $\tau_{\rm diff}$,
\begin{eqnarray}
\tau_{\rm diff} \sim \frac{r_c^2}{\alpha c_s^2 /\Omega_{\rm K}}
\sim 10^3 \left(\frac{\alpha}{10^{-3}}\right)^{-1} {\rm ~yr}.
\end{eqnarray}
The gap opening timescale itself would be comparable to $\tau_{\rm diff}$
\citep{sasaki_etal10}; therefore, both timescales are much shorter than
accretion and migration timescales of satellites.
In our calculations, the disk gas is abruptly dissipated with 
$\tau_{\rm diff}$, such that the satellites are ``frozen'' 
at that time.
Note that although hydrodynamic simulation suggests 
that the gap opening could not completely truncate the 
infalling gas (e.g., \citealt{lubow_etal99}), 
even in this situation, results that are shown in this study 
remain unchanged because the infall rate would be reduced
by several orders of magnitude. 
As long as the inflow cutoff timescale is shorter than accretion and migration
timescales, which are $\sim 10^5~{\rm years}$, the results are not affected.
That is, the results are not sensitive to the decay rate of the inflow.

As the location of the inner edge of the protosatellite disk is not well constrained,
in our calculations, we set the disk inner edge at 
$5~R_{\rm P}$, which is slightly outside the current corotation
radius of Jupiter ($\simeq 2.25~R_{\rm J}$).
At the disk inner edge, the gas surface density of the disk 
smoothly vanishes with a hyperbolic tangent function with 
the width $\Delta r$. We adopt $\Delta r = 0.2~R_{\rm P}$, which is
comparable to the disk scale height $H$.
In addition, satellites that migrate inside $4~R_{\rm P}$ are
disregarded from calculation mainly for computational reasons,
which is also described in Section~\ref{sec:tidal}.

\subsection{Solid Inflow Model}
\label{sec:solid_model}
Based on the model used by \citet{canup06},
solid bodies are added to the calculation between $r_{\rm in}= 5~R_{\rm P}$
and $r_c= 30~R_{\rm P}$ at a rate
$F_{\rm P}\eta_{\rm ice} f_{\rm d,in}/100 $, where $f_{\rm d,in}$ and $\eta_{\rm ice}$ are 
the scaling factors of the amount of solid material and the increase of solid materials due to 
ice condensation outside the ``ice line,'' respectively.
The solid infall flux per unit surface area is assumed to be uniform because 
we assume a uniform infall
flux of gas $F_{\rm P}$, which is stated in Section~\ref{sec:gas_density}.
When $\eta_{\rm ice} f_{\rm d,in}= 1$, the gas-to-solid ratio in the inflow is assumed to be 100. 
However, the ``effective''
gas-to-solid ratio in the disk is not fixed to 100, because gas surface density is 
also parameterized by $f_{\rm g}$ and the solid materials which are decoupled
from the gas evolve independently. (see Table~\ref{tbl:parameters} for the 
amount of gas of each run.)
We assume an exponential decay of the solid inflow with the timescale
of $\tau_{\rm dep}$.

\begin{deluxetable}{lccccc}
\tablecolumns{8}
\tablewidth{0pc}
\tablecaption{Parameters for Each Simulation}
\startdata
\hline \hline
Model (runs) & $f_{\rm g}$ & $f_{\rm d,in}$ & $C_{\rm I}$ & Ice Line\\
\hline
model 1(a,b,c) & 1 & 1 & 1 & No \\
model 2(a,b,c) & 1 & 1 & 1 & Yes \\
model 3(a,b,c) & 10 & 1 & 1 & No\\
model 4(a,b,c) & 1 & 0.5 & 1 & No \\
model 5(a,b,c) & 1 & 2 & 1 & No \\
model 6(a,b,c) & 1 & 1 & 0.1 & No \\
model 7(a,b,c) & 1 & 1 & 10 & No 
\enddata
\tablecomments{
Three runs referred to as a, b, and c (thus, a total of 21 runs) are performed for each model.
$f_{\rm g}$ and $f_{\rm d,in}$ are scaling factors for gas
surface density and the solid inflow, respectively. 
$C_{\rm I}$ is the scaling factor for the type I migration speed. For model~2,
solid enhancement outside the ice line is considered. 
}
\label{tbl:parameters}
\end{deluxetable}

Although the infall of solid components is the continuous dust infall in the gas flow,
it is mimicked by adding bodies with mass
 $M_{\rm add} = 5 \times 10^{-7} \eta_{\rm ice}(r/15 R_{\rm P})^3 M_{\rm P}$
 with a random azimuthal location at some time interval (see Supplementary 
 Information in \citealt{canup06}). Note that although a uniform solid inflow flux
 per area is assumed, radially dependent masses are used for adding bodies to avoid
 an imbalance in the number of adding bodies between radial zones.
 A total of 2,000-6,000 bodies are calculated for single runs, depending on the runs.
The initial velocity dispersion of the protosatellites is
set to be their escape velocity. The corresponding initial 
eccentricity $e$ and inclination $i$ are given by $v_{\rm esc}
= \sqrt{e^2 + i^2}v_{\rm K}$ with $e= 2i$ \citep{ida_makino92}.

In some runs, solid enhancement outside the ice line is
adopted. Assuming the condensation temperature of 
$\sim 160$ K, the location of the ice line is derived
from Equation~(\ref{eq:temp}), and we consider
that the ice line moves inward as the gas inflow rate decreases.
Because $T_d \propto F_{\rm P}^{1/4}$ and $T_d \propto r^{-3/4}$,
the location of the ice line is proportional to $F_{\rm P}^{1/3}$:
$r_{\rm ice} = 30 \exp(-t/3\tau_{\rm dep}) R_{\rm P}$.

\subsection{Orbital Integration}
The orbits of satellitesimals are calculated by numerically
integrating the equation of motion of the particle $k$ at $\textbf{\textit{r}}_k$
in planetocentric coordinates,
\begin{eqnarray}
\frac{d^2 \textbf{\textit{r}}_k}{dt^2}
& = & -GM_{\rm P} \frac{\textbf{\textit{r}}_k}{ |\textbf{\textit{r}}_k|^3} 
- \sum_{j \neq k} GM_j 
\frac{\textbf{\textit{r}}_k - \textbf{\textit{r}}_j}{|\textbf{\textit{r}}_k - \textbf{\textit{r}}_j|^3} 
- \sum_{j} GM_j \frac{\textbf{\textit{r}}_j}{ |\textbf{\textit{r}}_j|^3} 
\nonumber\\
&   &   
+ \textbf{\textit{F}}_{\rm damp} + \textbf{\textit{F}}_{\rm mig} + \textbf{\textit{F}}_{\rm tide},
\end{eqnarray}
where $k, j$ = 1, 2, ..., the first term on the right-hand side 
is the gravitational force of the central planet, the second term
is mutual gravity between the bodies, and the third is an indirect term. 
$\textbf{\textit{F}}_{\rm damp}$, $\textbf{\textit{F}}_{\rm mig}$, 
and $\textbf{\textit{F}}_{\rm tide}$ are specific forces due to
gravitational eccentricity damping, semimajor axis damping (type I migration),
and tidal eccentricity damping, respectively, which will be
explained in Section~\ref{sec:timescales}.

For numerical integration, we use the fourth-order Hermite scheme
\citep{makino_aarseth92} with a hierarchical individual time 
step \citep{makino91}.
When physical radii of two spherical bodies overlap, 
we consider the bodies to merge, conserving total mass and 
momentum and assuming perfect accretion.
The physical radius of a body is determined by its mass $M$
and internal density $\rho$ as
\begin{equation}
R = \left(\frac{3}{4 \pi}
\frac{M}{\rho}
\right)^{1/3},
\end{equation}
where we adopt $\rho = 3~{\rm g~cm}^{-3}$.

\subsection{Characteristic Timescales}
\label{sec:timescales}
\subsubsection{Gravitational Interaction with Disk Gas}
We consider damping of orbital eccentricity, inclination,
and semimajor axis due to disk-satellite interactions.
A satellite gravitationally perturbs the disk gas
and excites density waves, which damp $e, i,$ and $a$
of the satellite\footnote{This damping is sometimes called ``tidal damping.''
Because damping of orbital elements through tidal energy dissipation within planets
and tidal interactions between planets and satellites through their tidal deformation
are also considered in this study, we use the term ``tidal''
only when we refer to the tidal energy dissipation in planets and the tidal interaction 
between planets and satellites.} 
(e.g., \citealt{goldreich80}; \citealt{ward86}; 
\citealt{artymowicz93}).

The force formulae for $e$-damping and $i$-damping 
($\textbf{\textit{F}}_{\rm damp}$) are 
the same as that in Equations~(2)-(4) in \citet{ogihara_etal10},
and the $e$-damping timescale is \citep{tanaka04}
\begin{eqnarray}
t_e &=& -\frac{e}{\dot{e}}  = 
\frac{1}{0.78}\left(\frac{M}{M_{\rm P}}\right)^{-1} 
\left(\frac{\Sigma_{\rm g} r^2}{M_{\rm P}}\right)^{-1}
\left(\frac{c_s}{v_{\rm K}}\right)^{4} \Omega^{-1},\\
&=& 4.1 \times 10^3 f_{\rm g}^{-1}
\left(\frac{M}{10^{-4}M_{\rm P}}\right)^{-1}
\left(\frac{M_{\rm P}}{M_{\rm J}}\right)^{-3/2}
\left(\frac{r}{20R_{\rm P}}\right)^{3/4}
\nonumber \\
&& \times 
\left(\frac{R_{\rm P}}{R_{\rm J}}\right)^{3/4}
\left(\frac{\tau_{\rm G}}{5\times 10^6 {\rm ~yr}}\right)^{-1/2}
{\rm ~yr}.
\label{eq:Tanaka_e_s}
\end{eqnarray}
The force formula for $a$-damping ($\textbf{\textit{F}}_{\rm mig}$) is 
the same as that in Equation~(3) in \citet{ogihara_etal10}, and the 
migration timescale is \citep{tanaka_etal02}
\begin{eqnarray}
t_a &=& - \frac{a}{\dot{a}} = \frac{1}{C_{\rm I}}
\frac{1}{2.7+1.1q} \left(\frac{M}{M_{\rm P}}\right)^{-1}
\left(\frac{\Sigma_{\rm g} r^2}{M_{\rm P}}\right)^{-1}
\left(\frac{c_s}{v_{\rm K}}\right)^{2} \Omega^{-1},\\
&=& 1.2 \times 10^5 C_{\rm I}^{-1} f_{\rm g}^{-1}
\left(\frac{M}{10^{-4}M_{\rm P}}\right)^{-1}
\left(\frac{M_{\rm P}}{M_{\rm J}}\right)^{-1}
\left(\frac{r}{20R_{\rm P}}\right)^{1/2}
\nonumber \\
&& \times 
\left(\frac{R_{\rm P}}{R_{\rm J}}\right)^{1/2}
\left(\frac{\tau_{\rm G}}{5\times 10^6 {\rm ~yr}}\right)^{-1/4}
{\rm ~yr},
\label{eq:Tanaka_a_s}
\end{eqnarray}
where $\Sigma_{\rm g} \propto r^{-q}$ and $q= 3/4$ is adopted
from Equation~(\ref{eq:surface_density}).
$C_{\rm I}$ is a scaling factor to allow for retardation and acceleration of type I migration.
We include neither reverse torque of type I migration near the disk
edge \citep{masset_etal06} nor the radiative effect through the entropy gradient
\citep{paardekooper_etal10}; such issues will be resolved in future studies.
Instead, we handle $t_a$ as a parameter in 
some runs and focus on the effect of the eccentricity trap, which is 
a physical mechanism to halt type I migration near the disk inner edge
\citep{ogihara_etal10}. 
This trapping arises when a body with an elliptical orbit straddles the disk inner edge;
the body gains a net positive torque from the gas disk, which can compensate
for negative type I migration torques that are exerted on a resonantly interacting 
convoy of bodies.

As discussed in \citet{ogihara_etal10}, \citet{tanaka04} assumed small
$e$ and $i$ ($\ll c_s/v_{\rm K}$) to obtain the damping
rate of eccentricity. \citet{ostriker99}, \citet{papaloizou00}, and
\citet{muto_etal11} 
showed that the formulae of $e$-damping and $a$-damping must be multiplied 
by correction factors in the case of supersonic flow. When the correction 
factors are added to $t_e$ and $t_a$, it is suggested that the eccentricity trap 
becomes more efficient because the correction factor for $t_a$ is a stronger
function of $ev_{\rm K}/c_s$ than that for $t_e$. 
However, we neglect these factors for simplicity.

\subsubsection{Tidal Dissipation}
\label{sec:tidal}
Tidal dissipation within satellites plays an important role in
satellite evolution. Thus, we incorporate tidal eccentricity damping 
of satellites as a force acting on them \citep{papaloizou10}:
\begin{equation}
\textbf{\textit{F}}_{\rm tide} = -\frac{2}{t_{e,{\rm tide}}}
\frac{(\textbf{\textit{v}} \cdot \textbf{\textit{r}})\textbf{\textit{r}}}{|\textbf{\textit{r}}|^2},
\end{equation}
where the damping timescale is \citep{goldreich_soter66}
\begin{eqnarray}
t_{e,{\rm tide}}&=&-\frac{e}{\dot{e}}
= \frac{4}{63} \frac{M}{M_{\rm P}}
\left(\frac{a}{R}\right)^{5}
\tilde{\mu}Q \Omega^{-1}\\
&\simeq&  \frac{4}{63} \left(\frac{M}{M_{\rm P}}\right)^{-2/3}
\left(\frac{a}{R_{\rm P}}\right)^{13/2}
\left(\frac{\rho}{\rho_{\rm P}}\right)^{5/3}
\tilde{\mu}Q \Omega^{-1}\\
&\sim& 10^9 
\left(\frac{M}{10^{-4}M_{\rm P}}\right)^{-2/3}
\left(\frac{a}{20R_{\rm P}}\right)^{13/2}
\left(\frac{\tilde{\mu}Q}{1000}\right)
\nonumber \\ && \times
\left(\frac{R_{\rm P}}{R_{\rm J}}\right)^{3/2}
\left(\frac{M_{\rm P}}{M_{\rm J}}\right)^{-1/2}
{\rm ~yr}.
\label{eq:e-damp_tide}
\end{eqnarray}
Here, $\tilde{\mu}$ ($\simeq 3/2k_2;$ $k_2$ is the love number) and $Q$ are 
the effective rigidity and
the tidal dissipation function of the satellites, respectively.
Although there is an uncertainty in $Q$, it is likely that both 
$Q$ and $\tilde{\mu}$ are of the order of 10-100 \citep{yoder95}.
We thus use $Q' \equiv \tilde{\mu}Q = 1000$ for our \textit{N}-body simulations.

Note that tidal torque on the satellite changes its 
semimajor axis. 
The migration timescale is \citep{goldreich_soter66}
\begin{eqnarray}
t_{a,{\rm tide}} &=& \left|\frac{a}{\dot{a}}\right|
=\frac{Q_{\rm P}}{3 k_{2,{\rm P}}}
\left(\frac{M}{M_{\rm P}}\right)^{-1}
\left(\frac{a}{R_{\rm P}}\right)^{5} \Omega^{-1}\\
&\sim& 10^{13}  \left(\frac{Q_{\rm P}}{10^5}\right)
\left(\frac{k_{2,{\rm P}}}{0.5}\right)^{-1}
\left(\frac{M}{10^{-4}M_{\rm P}}\right)^{-1} 
\nonumber \\ && \times
\left(\frac{a}{20 R_{\rm P}}\right)^{13/2}
\left(\frac{R_{\rm P}}{R_{\rm J}}\right)^{3/2}
\left(\frac{M_{\rm P}}{M_{\rm J}}\right)^{-1/2}
{\rm ~yr},
\end{eqnarray}
where $Q_{\rm P}$ and $k_{2,{\rm P}}$ are the 
tidal dissipation function and the love number of the 
central planet, respectively. 
The direction of migration is determined by the position of the satellite with 
respect to the corotation radius with the spin of the host planet.
For example, if the satellite is on the inside of
the corotation radius, the satellite moves inward. 
Because the migration timescale is generally much longer than the simulation time,
we neglect this effect. 
The change in semimajor axis due to dissipation within the satellite is also
ignored because this timescale is also long except for very close-in satellites.
The migration timescale has a strong dependence on the semimajor axis;
thus, extremely close satellites inside the corotation radius would fall onto 
the planet in a short time.
As previously described, the current corotation radius of Jupiter is $\simeq 2.25~R_{\rm P}$;
satellites that migrate inside of this radius would be quickly lost to the planet.
Because of both computational cost and uncertainty in the corotation radius,
we disregard bodies that reach $4~R_{\rm P}$ from our calculation.

\section{TYPICAL RESULTS OF \textit{N}-BODY SIMULATIONS}
\label{sec:n-body}
In this section, we show typical results with our fiducial parameters;
the parameters for each simulation are listed in Table~\ref{tbl:parameters}.
First, we perform \textit{N}-body simulations in a gas disk that dissipates with
the global dissipation timescale of $\tau_{\rm dep} = 10^6~{\rm years}$,
assuming that the solid inflow flux also decreases with the same
timescale. Then, the host planet opens a gap in the circumstellar disk
at a certain time to enable the gas and solid inflow fluxes to rapidly 
decrease with a disk viscous diffusion timescale of $\tau_{\rm diff}
\sim~10^3~{\rm years}$. However, because the timing cannot be determined,
we randomly choose the moment and perform the subsequent simulation.
We mainly focus on the \textit{N}-body results in a gas disk because we
found that orbital configurations do not change after
rapid gas depletion.

\subsection{Growth and Migration before Gap Opening}
\label{sec:growth}
Orbital evolution of satellites embedded in a gas disk is presented in this subsection.
In our fiducial runs (models 1a, 1b, and 1c), 
the gas scaling factor $f_{\rm g}= 1$ is used, the 
solid inflow rate is fixed at $f_{\rm d,in}= 1$, 
and the migration rate $C_{\rm I}= 1$ is assumed to be that predicted by the linear 
theory. We note that these parameters are not
unique values but contain some uncertainties. In Section~\ref{sec:parameters},
we will discuss the dependency of the results on these parameters.
Other effects such as the supersonic correction factor of gravitational damping
formulae and solid enhancement outside the ice line are not
considered.

\begin{figure}[htbp]
\epsscale{1.2}
\plotone{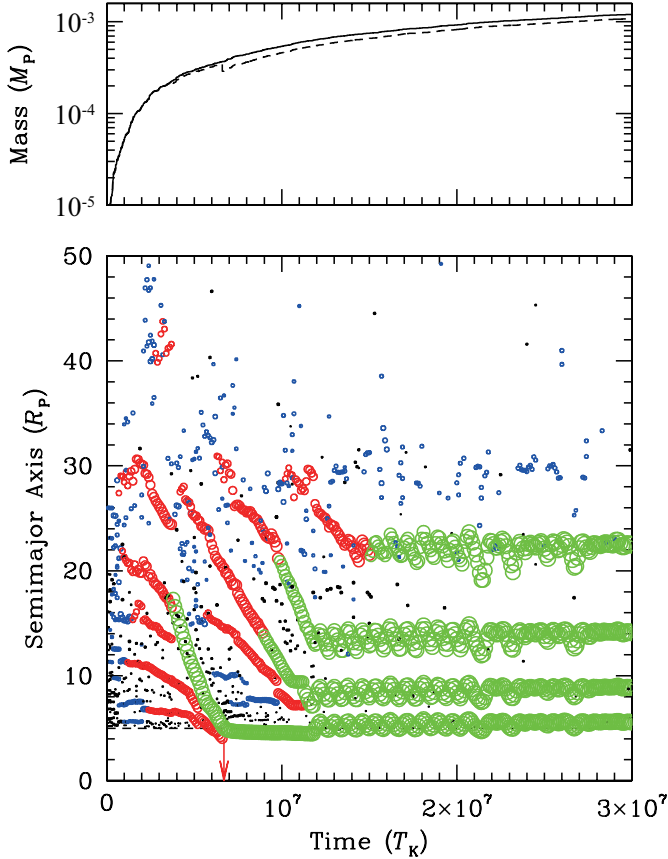}
\caption{
Top: Evolution of the solid mass for model~1a. The dashed line shows
the total mass added to the calculation and the solid line shows the mass
remaining within the disk.
Bottom: Orbital evolution of the satellites for model~1a. For each time,
30 most massive satellites are plotted. The circles represent bodies, and
the radii of the circles are proportional to the physical radii of the bodies.
In the electronic version, bodies with $M>10^{-6} M_{\rm P}$, 
$M>10^{-5} M_{\rm P}$, and $M>10^{-4} M_{\rm P}$ are expressed with 
blue, red, and green circles, respectively. The arrow indicates that the 
satellite ($\geq 10^{-5}~M_{\rm P}$) migrates inside $4~R_{\rm P}$.
}
\label{fig:t_a_run1}
\end{figure}

The orbital evolution for model~1a is shown in the bottom panel of Figure~\ref{fig:t_a_run1}.
In this figure, 30 most massive satellites at each time are plotted as
circles, the radii of which are proportional to the radii of the bodies. 
$T_{\rm K} (\simeq 0.03~{\rm years})$ is the orbital period at $a= 20~R_{\rm J}$ around a Jovian
mass planet ($M= M_{\rm J}$). The dashed line at $5~R_{\rm P}$ represents
the location of the disk inner edge.
When a satellite with mass greater than $10^{-5}~M_{\rm P}$ falls onto
the host planet, it migrates inside $4~R_{\rm P}$; the descent
is indicated by an arrow. The evolution of mass is shown in the top panel
of Figure~\ref{fig:t_a_run1}, which can be useful for physical understanding of what we
observe here. The dashed line refers to the total mass added, and the solid line to the
solid mass remaining within the disk.

Although many satellitesimals are not initially placed,
solid materials are continuously added to the calculation, and
the growth of satellitesimals proceeds in a manner similar to that of 
planetesimals around a star (e.g., \citealt{kokubo_ida98}). 
Through orbital repulsion of oligarchs that form from planetesimals,
their orbital separations are kept wider than five mutual Hill radii. 
The typical orbital separation is $\Delta r \simeq 10-15~r_{\rm H}$, where
$r_{\rm H}$ is the mutual Hill radius of oligarchs.
As is discussed in \citet{canup06}, it is important to note that owing to the low solid inflow rate, 
the growth rate of a satellite is governed by the solid inflow flux. 
The accretion timescale is given by
\begin{eqnarray}
\tau_{\rm acc,inflow} 
&\simeq& 1.8 \times 10^5 \eta_{\rm ice}^{-1}
f_{\rm d,in}^{-1} 
\left(\frac{r}{20R_{\rm P}}\right)^{-2}
\nonumber \\ && \times
\left(\frac{M}{10^{-4}M_{\rm P}}\right)^{2/3}~{\rm yr}.
\label{eq:acc}
\end{eqnarray}
More detailed descriptions of satellite growth and the growth timescale
are presented in Appendix~\ref{app:growth}.

When the satellite grows to a critical mass, which is determined by a 
balance between the accretion time $\tau_{\rm acc,inflow}$ and 
migration time $t_a$, it begins to undergo orbital decay. 
The critical mass $M_{\rm crit}$ is estimated from 
Equations~(\ref{eq:Tanaka_a_s}) and (\ref{eq:acc}) as
\begin{eqnarray}
M_{\rm crit} &\simeq&
8 \times 10^{-5}
\left(\frac{\eta_{\rm ice} f_{\rm d,in}}{C_{\rm I} f_{\rm g}}\right)^{3/5} 
\left(\frac{M_{\rm P}}{M_{\rm J}}\right)^{-3/5}
\left(\frac{r}{20R_{\rm P}}\right)^{3/2}\nonumber \\
&& \times \left(\frac{R_{\rm P}}{R_{\rm J}}\right)^{3/10}
\left(\frac{\tau_{\rm G}}{5\times 10^6~{\rm yr}}\right)^{-3/20}~M_{\rm P}.
\label{eq:m_crit}
\end{eqnarray}
The critical mass in our fiducial case is 
$\sim 8 \times 10^{-5}~M_{\rm P}$, which is comparable to the masses of
of the Galilean satellites. Because satellites grow to some extent
during migration, the factor ($\eta_{\rm ice} f_{\rm d,in}/C_{\rm I} f_{\rm g}$)
in Equation~(\ref{eq:m_crit}), which places a lower limit on the satellite mass,
would be smaller than unity for the case of the Jovian satellite formation.
A discussion of parameter dependence is given in 
Section~\ref{sec:parameters}.

The migrating satellites exhibit dynamics different from those
in \citet{canup06} because of the existence of the disk inner cavity.
The satellite that reaches the disk inner edge no longer
experiences gas drag, and migration ceases.
The subsequently migrating satellite is captured into a 2:1 mean motion
resonance with the inner satellite, leading to excitation of eccentricities 
($\simeq 0.1$) of both bodies. 
The inner satellite that resides near the disk inner edge gains
a positive torque from the disk, and eventually, the satellites captured into 
the 2:1 mean motion resonance remain in the disk.
This trapping mechanism of bodies near the disk edge has been
both numerically and analytically investigated by \citet{ogihara_etal10}.
They refer to the positive torque gained by the satellite and the 
trapping mechanism as  the edge torque and the eccentricity trap, respectively.
It should be emphasized that we do not establish a rigid wall
at the disk inner edge or manually impose an edge torque to the satellite at the edge.
We apply only instantaneous drag forces that have been physically examined
in previous studies \citep{tanaka04} onto all bodies.
If the body at the disk edge has a nonzero eccentricity, it moves back and forth
between the inner cavity and the gas disk.
When it moves into the gas disk near the apocenter, the tangential velocity of the body
is slower than the local gas velocity; therefore, the body suffers a tailwind.
On the other hand, when it enters into the cavity near the pericenter, where the tangential 
velocity of the body is faster than the local circular Keplerian velocity, the body
does not feel gas drag, which results in a net positive torque on the body.
\citet{ogihara_etal10} determined that if the edge torque balances with the negative type I 
torque exerted on the bodies, the eccentricity trap works effectively and
other migrating bodies can even be trapped without pushing the innermost body into the
cavity.
Figure~\ref{fig:t_a_run1} shows that the eccentricity trap 
is retained, except at $t\simeq 7 \times 10^6~T_{\rm K}$.
In fact, this trapping ordinarily also occurs in other simulations  
not shown here. 
However, an upper limit exists on the total number of satellites
that can be retained in the disk, which is discussed in Section~\ref{sec:number}. 

\begin{figure}[htbp]
\epsscale{1.2}
\plotone{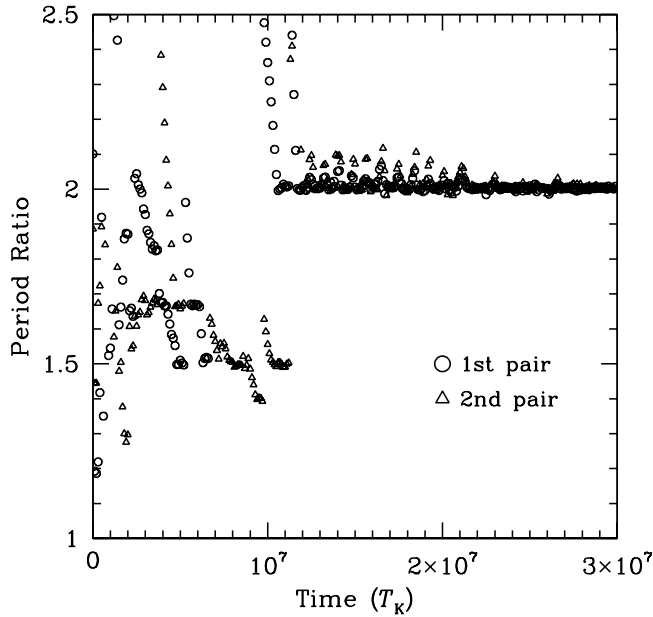}
\caption{Evolution of the period ratios for model~1a.
Circles and triangles express period ratios of the innermost
and the second innermost pairs, respectively.}
\label{fig:t_r2_run1}
\end{figure}

We also found that the 2:1 mean motion commensurability is 
universal in our \textit{N}-body simulations. Figure~\ref{fig:t_r2_run1}
shows the evolution of period ratios. Open circles and triangles 
represent period ratios of the innermost and the second innermost
pairs, respectively. Although some satellites are pushed inside 
the 2:1 resonant location by outer migrating satellites and are captured
into closer first-order resonances (e.g., 3:2 and 4:3), almost all 
pairs, whose period ratios decrease above two,
are locked in the 2:1 resonance. Hence, the period ratio is two.
However, the resonant angles do not always librate 
about fixed values but sometimes exhibit circulation.
The commensurate value and the occurrence of the eccentricity
trap, which regulates the number of trapped bodies outside the disk
edge, depends on eccentricity damping and migration rates, which is 
discussed in Section~\ref{sec:parameters}.

Satellites with mass $M_{\rm crit}$ begin to move inward
and continue to grow during migration even after they are captured into
a resonance. At $3\times 10^7~T_{\rm K}$ when we stopped calculations,
the masses were $2.4 \times 10^{-4}~M_{\rm P}$ (innermost),  
$2.5 \times 10^{-4}~M_{\rm P}$ (second innermost),
$2.9 \times 10^{-4}~M_{\rm P}$ (third innermost), and 
$3.0 \times 10^{-4}~M_{\rm P}$ (fourth innermost), which are heavier than
the current Galilean satellites. For satellites with masses comparable
to those of the Galilean satellites to be formed, the gap opening time, in which 
the solid inflow is terminated, should be earlier than $3\times 10^7~{\rm T_{\rm K}}$. 
In addition, the critical mass for migration 
(Equation~(\ref{eq:m_crit})) must not exceed
the masses of the Galilean satellites.
A discussion of satellite mass appears in Section~\ref{sec:mass}.

\begin{figure}[htbp]
\epsscale{1.2}
\plotone{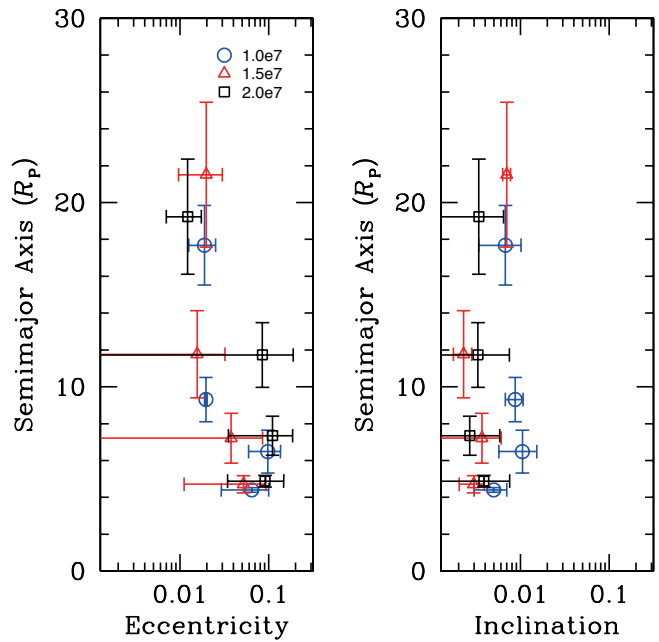}
\caption{Averaged semimajor axes, eccentricities, and inclinations of 
the four innermost satellites at $t= 1 \times 10^7~T_{\rm K}$ (blue circles),
$t= 1.5 \times 10^7~T_{\rm K}$ (red triangles), and
$t= 2 \times 10^7~T_{\rm K}$ (black squares) for models 1a, 1b, and 1c, respectively.
Error bars denote the 1$\sigma$ dispersion. 
(A color version of this figure is available in the online journal.)
}
\label{fig:e_a}
\end{figure}

We performed two additional runs
(models 1b and 1c) under the same conditions but with different
random numbers for additional satellitesimal locations,
and we confirmed that the overall evolutionary tendency 
is identical.
The averaged orbital distributions of the four innermost satellites 
($\geq 10^{-5} ~M_{\rm P}$) at each time
are plotted in Figure~\ref{fig:e_a}.
Circles, triangles, and squares represent averaged values
over three simulations at $1 \times 10^7$, 
$1.5 \times 10^7$, and $2 \times 10^7~T_{\rm K}$, 
respectively.
Error bars indicate 1$\sigma$ dispersion.
The semimajor axis distributions that are obtained at given times are
similar to those of the Galilean satellites. 
The three inner satellites are usually in the 2:1 resonance, and the fourth
one is often captured into the 2:1 resonance. We set the disk inner 
edge at $5~R_{\rm P}$ so that the innermost one is located at $5~R_{\rm P}$.
This orbit is slightly interior compared with that of Io; thus, satellites
in Figure~\ref{fig:e_a} generally exist closer than the Galilean moons.
The inclinations are kept small compared to the excited eccentricities
due to resonant interactions, and they are
comparable to the current inclinations of the Galilean satellites.
Table~\ref{tbl:galilean_satellites} shows $e$ and $i$ of the Galilean satellites.

\subsection{Orbital Evolution after Gap Opening}
\label{sec:continue}
The inflow is abruptly cut off because of the gap opening; thus, the 
disk gas is depleted at a certain time during the evolution in 
Figure~\ref{fig:t_a_run1}. Because the time of 
the gap opening is unknown, we randomly choose the moment and 
continue calculations with gas dissipation on the timescale of
$\tau_{\rm diff} = 10^3~{\rm years} \simeq 3 \times 10^4~T_{\rm K}$.

\begin{figure}[htbp]
\epsscale{1.2}
\plotone{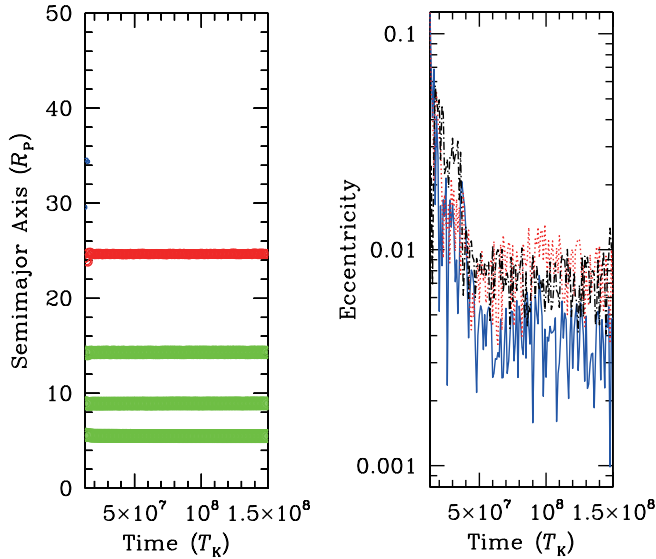}
\caption{Subsequent evolution of model~1a with rapid gas dissipation.
Left: Evolution of semimajor axis. Right: Evolution of eccentricities of
the three inner satellites. The eccentricities of the innermost,
the second innermost, and the third innermost satellites are
expressed with the solid red line, the dotted blue line, and the dashed
black line, respectively. 
(A color version of this figure is available in the online journal.)
}
\label{fig:continue}
\end{figure}

Figure~\ref{fig:continue} shows subsequent evolution of model~1a 
in which the orbital calculation is restarted at $1.3\times 10^7~T_{\rm K}$.
Considering the formation of the Galilean moons in which the fourth satellite, Callisto,
is not in a mean motion resonance, it is likely that this satellite underwent
little or no migration. 
To achieve little significant migration, the gap opening timing, after which the
circumplanetary disk dissipates on the timescale of $\tau_{\rm diff}$, must be
comparable to the timing of the
completion of Callisto formation. We assume the time to be $1.3\times 10^7~T_{\rm K}$.
An alternative scenario is that Callisto was accreted 
from the slowly inflowing materials during an imperfect gap opening \citep{sasaki_etal10},
which naturally explains the inferred undifferentiated interior of Callisto \citep{barr08}.
Both scenarios explain the non-resonant Callisto.
However, because the timing of gap opening cannot theoretically be predicted,
more detailed discussion of characteristics of Callisto is left for future study.

The left panel of Figure~\ref{fig:continue} shows the semimajor axis evolution.
Because the satellites are in mean motion resonances and their
orbital separations are wide ($\simeq 15~r_{\rm H}$),
orbital configurations hardly change.
Although the semimajor axis can be slightly decreased because of
tidal dissipation, such an effect is negligible. Tidal torque can
also change the semimajor axis; however, we ignore this effect, as previously stated.
The right panel shows the eccentricity evolutions of the innermost
(solid line), the second innermost (dotted line), and third innermost
(dashed line) satellites. 
The $e$-damping timescale 
of the innermost satellite, which is located at $5~R_{\rm P}$, is estimated as
$\sim 10^6~{T_{\rm K}}$ (Equation~(\ref{eq:e-damp_tide})). 
This result is not strictly consistent with the actual damping time ($\sim 10^7~{T_{\rm K}}$)
of the innermost satellite observed in Figure~\ref{fig:continue}, 
because eccentricities of outer satellites in resonances are also dragged down 
by tidal dissipation in the innermost satellite via resonant interactions between them. 
Thus, the eccentricities of the resonant satellites become $\lesssim 0.01$,
which is consistent with thos of the Galilean satellites (Table~\ref{tbl:galilean_satellites}).
Regarding the evolution of the resonant angles, as the orbital energy is 
dissipated because of tidal $e$-damping, the satellites evolve deeper into
the resonances, leading to a small amplitude libration around $0^\circ$ or
$180^\circ$. 
Again, not all of resonant angles librate about fixed values.
In three runs of all fifteen simulations (models 1, 3-6), the Laplace relationship,
which indicates that $\theta_5$ librates, can be observed.
 The Galilean-like configuration ($\theta_1$,
$\theta_2$, $\theta_3$, and $\theta_5$ librate while $\theta_4$ 
circulates) is reproduced by one run of all. 
The definition of the resonant angle is presented in 
Appendix~\ref{app:resonant}.

\subsection{Origin of Compositional Gradient}
By tracking the orbital evolution of all the satellites in the \textit{N}-body simulation,
satellite composition can be discussed.
As mentioned in Section~\ref{sec:intro}, a compositional gradient exists
in the Galilean satellites. That is, bulk densities of Io, Europa,
Ganymede, and Callisto are 3.53, 2.99,
1.94, and 1.83~${\rm g~cm^{-3}}$, respectively.
The composition of satellites is considered to be a reflection of disk temperature;
thus, we discuss in this subsection the origin of the compositional gradient in terms of
disk temperature. Disks can be classified into three classes:
(i) hot, in which only rocky materials exist;
(ii) cool, in which the ice condensation line is located in the satellite-forming
region, and rocky and icy objects coexist;
(iii) cold, in which water vapor can condense into ice in all parts of the disk.

\begin{figure}[htbp]
\epsscale{1.2}
\plotone{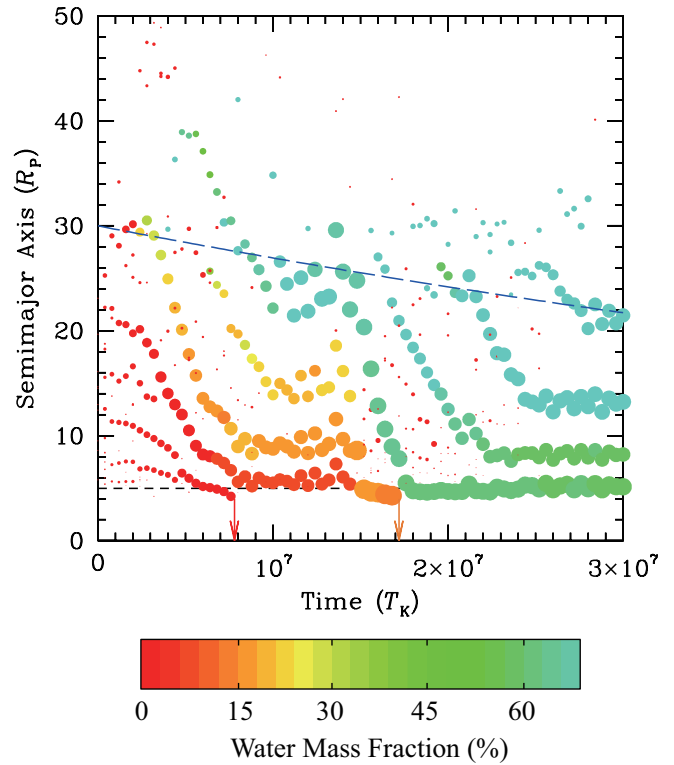}
\caption{Evolution of semimajor axis and water mass fraction for model~2a.
The color of each object corresponds to its water mass fraction.
The long dashed line represents the location of the ice line $r_{\rm ice}$.
}
\label{fig:t_comp}
\end{figure}

To observe the effect of the ice line, we performed an additional set of
\textit{N}-body simulations (model~2) by considering the enhancement of solid materials
outside the ice line assuming $\eta_{\rm ice}= 3$, and we track the accumulation 
and migration of satellites.
As noted in Section~\ref{sec:solid_model}, we assume that the ice line moves 
inward according to $r_{\rm ice} = 30 \exp(-t/3\tau_{\rm dep}) R_{\rm P}$,
which indicates that a cool disk is considered.
Figure~\ref{fig:t_comp} shows the evolution of model~2a, in which the colors
corresponds to the water (ice) mass fraction of satellites, from red (dry)
to light blue (up to 67\%). (The colors are visible in the online version of the paper.)
The long dashed line denotes the location of the ice line.
The water content is calculated
using the following prescription for components of satellites that
originated at $r$:
\begin{eqnarray}
\frac{M_{\rm water}}{M} = \frac{\eta_{\rm ice}-1}{\eta_{\rm ice}}
\simeq \left\{ \begin{array}{ll}
0 & (r\leq r_{\rm ice}) \\
0.67 & (r>r_{\rm ice}). \\
\end{array} \right.
\end{eqnarray}
In the case without the disk inner edge, icy materials 
can reach the innermost region with the timescale of type I migration
($\sim 10^6~T_{\rm K}$), and fine-tuning of timing is required for the
final survival of both rocky and icy satellites. 
The type I decay time should be comparable to the migration time of the ice 
line in the final state. 
However, as shown in Figure~\ref{fig:t_comp},
icy satellites stay outer region ($t = 1-1.5 \times 10^7~T_{\rm K}$).
Because satellites are lined up outside the disk inner edge, migrating icy satellitesimals
are consumed by the outer satellites captured into mean motion resonances,
leading to the deficiency of icy materials in the inner satellites.
Thus, contamination of inner satellites by
icy materials is prevented owing to the disk inner edge, and we
do not have to rely on the fine-tuning of the timing.
We find a monotonic increase in the water
content with an increase in the distance from the planet. 
Between $8\times 10^6$ and $1.5\times 10^7~T_{\rm K}$, the 
water mass fraction of the innermost
satellite is $\simeq 5\%$, while that of the outer satellite is $\simeq 60\%$.
This configuration is approximately consistent with the those of Galilean satellites.
However, after $1.8\times 10^7~T_{\rm K}$, the rocky inner satellites are
lost to the planet because of the violation of the eccentricity trap. As a result,
the surviving satellites contain a large quantity of ice ($\gtrsim 40\%$).

Rather than the effect of the ice condensation line, we find 
in a companion paper (Nimmo, Ogihara \& Ida, in preparation) that
the impact erosion of ices during accretion can produce the 
compositional gradient even in a cold disk. 
We find that the impact velocity increases with a decrease in the 
radial distance and exceeds $10~{\rm km~s^{-1}}$ at the orbits of
Io and Europa.
This observation is attributed to the high Keplerian orbital velocity ($\propto r^{-1/2}$)
and a large velocity dispersion.
Thus, nearly all ice in the innermost body, which corresponds to Io,
and most of the ice in the second innermost body, which corresponds to Europa,
evaporates through collisions. The amount of ice vaporization
depends on the
mass ratio of the target and the impactor, the impact velocity, and the
impact angle (\citealt{kraus_etal11}; \citealt{nimmo_korycansky12}).
Detailed analysis is presented in the companion paper, in which 
the evaporation of ice owing to tidal heating is also discussed.

To summarize, to reproduce the compositional gradient by
considering ice condensation outside the ice line, which indicates
a cool disk, the following two conditions are necessary. First,
the location of the ice line, in which there is an uncertainty, should be
sufficiently far to enable the formation of a large amount of
rocky satellitesimals. Second, once icy materials move
into the inner satellite formation region, the eccentricity trap of a resonant
convoy should not be broken. We also find that fine-tuning of timings is not required.
It is suggested that the gap opening at which the satellites cease to form
occurs during the period when satellites are trapped by the disk edge.
This inferred gap opening time is consistent with  
that assumed in Section~\ref{sec:continue}. 
In addition, we find that even if satellites
are formed in the cold disk, the origin of the compositional gradient can be 
explained by collisional erosion of volatiles, as discussed in the companion 
paper. 
Furthermore, we can exclude the possibility that the Galilean satellites are
formed in the hot disk, that is, Jupiter did not open the gap in the protoplanetary
disk when the circumplanetary disk was hot.

\section{DEPENDENCE ON PARAMETERS}
\label{sec:parameters}
In the previous section, we presented \textit{N}-body results of satellite accretion
for our fiducial parameters and reproduced the properties of the Galilean satellites 
(e.g., 2:1 commensurability). There are, however, uncertainties
in the parameters that we used. 
\begin{quote}
\textit{Gas surface density ($f_{\rm g}$)}: Following \citet{canup06} and
\citet{sasaki_etal10}, we assumed $\alpha = 5 \times 10^{-3}$ and 
$\tau_{\rm G} = 5\times 10^6~{\rm years}$,
leading to a gas-starved disk with $f_{\rm g}= 1$ (Equation~(\ref{eq:surface_density})). 
This assumption of a small mass disk
may be reasonable in the final stage of satellite formation because the 
inflow rate somehow decays, and gas surface density decreases.
Recently, \citet{fujii_etal11} suggested a low magnetic Reynolds number
and thus a low turbulent
viscosity, which may suggest a more massive disk ($f_{\rm g}>1$ in 
Equation~(\ref{eq:surface_density})).
\end{quote}
\begin{quote}
\textit{Amount of solid materials ($f_{\rm d,in}$)}: We assumed 
$f_{\rm d,in}= 1$ for the amount of solid materials, in which the gas-to-dust
ratio in the inflowing gas is considered to be 100 according to solar
metallicity. 
It is inferred that ``metal'' abundance in Jupiter is 
high (e.g., \citealt{saumon_guillot04}); thus, the inflowing gas would be
more metal-rich (thus $f_{\rm d,in}>1$).
This could be due to depletion of local gas by accretion onto Jupiter or that
toward the Sun, 
leaving an excess of solid materials. However, dust grain abundance may also
be reduced due to solids having been incorporated into planetesimals, which
decreases $f_{\rm d,in}$. Thus, depending on which effect is dominated,
$f_{\rm d,in}$ can be larger or smaller than unity.
\end{quote}
\begin{quote}
\textit{Efficiency of type I migration ($C_{\rm 1}$)}: In the fiducial model, the
type I migration speed is set to the value derived from the linear calculation
($C_{\rm 1}= 1$). 
Several mechanisms for reducing migration efficiency have been discussed
(e.g., \citealt{masset_etal06}; \citealt{paardekooper_etal10}). In fact, 
to reproduce the observed distributions of exoplanets,
$C_{\rm I}\lesssim 0.1$ \citep{ida_lin08}.
\end{quote}
In this section, therefore, we first examine the mechanism through which
the \textit{N}-body results are altered by adopting
different parameter values, and we attempt to link the model parameters to
the physical properties of formed satellites, such as the resonant relationship, 
by using semianalytical arguments.
This approach, in which dependences of parameters on properties of satellites
are investigated, is valid not only for
setting limits on parameters that reproduce the Galilean satellites
but also for discussing satellite formation in general, including satellite formation
around extrasolar giant planets.

\subsection{\textit{N}-body Simulations with Various Parameters}
\begin{figure}[htbp]
\epsscale{1.0}
\plotone{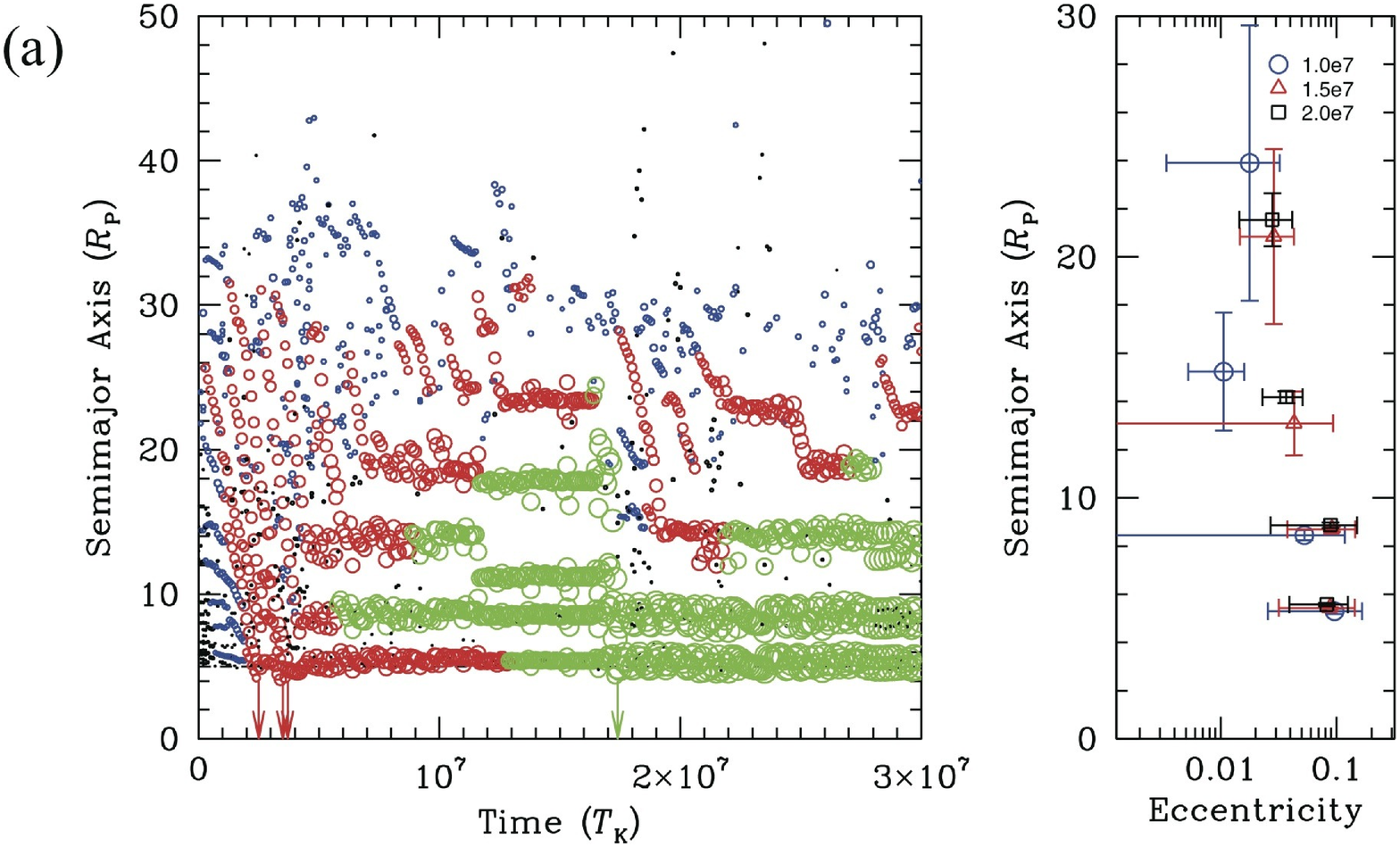}
\plotone{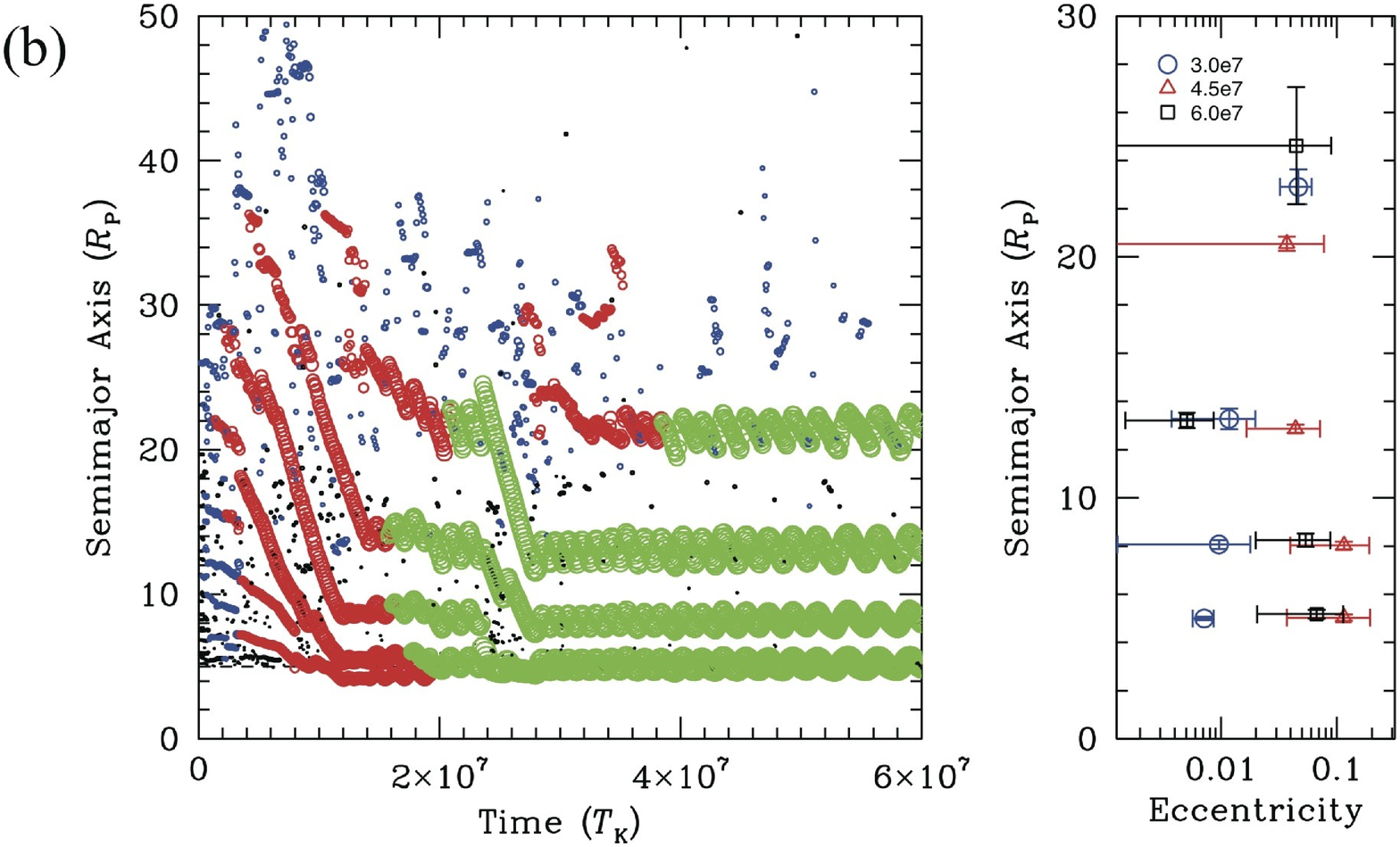}
\plotone{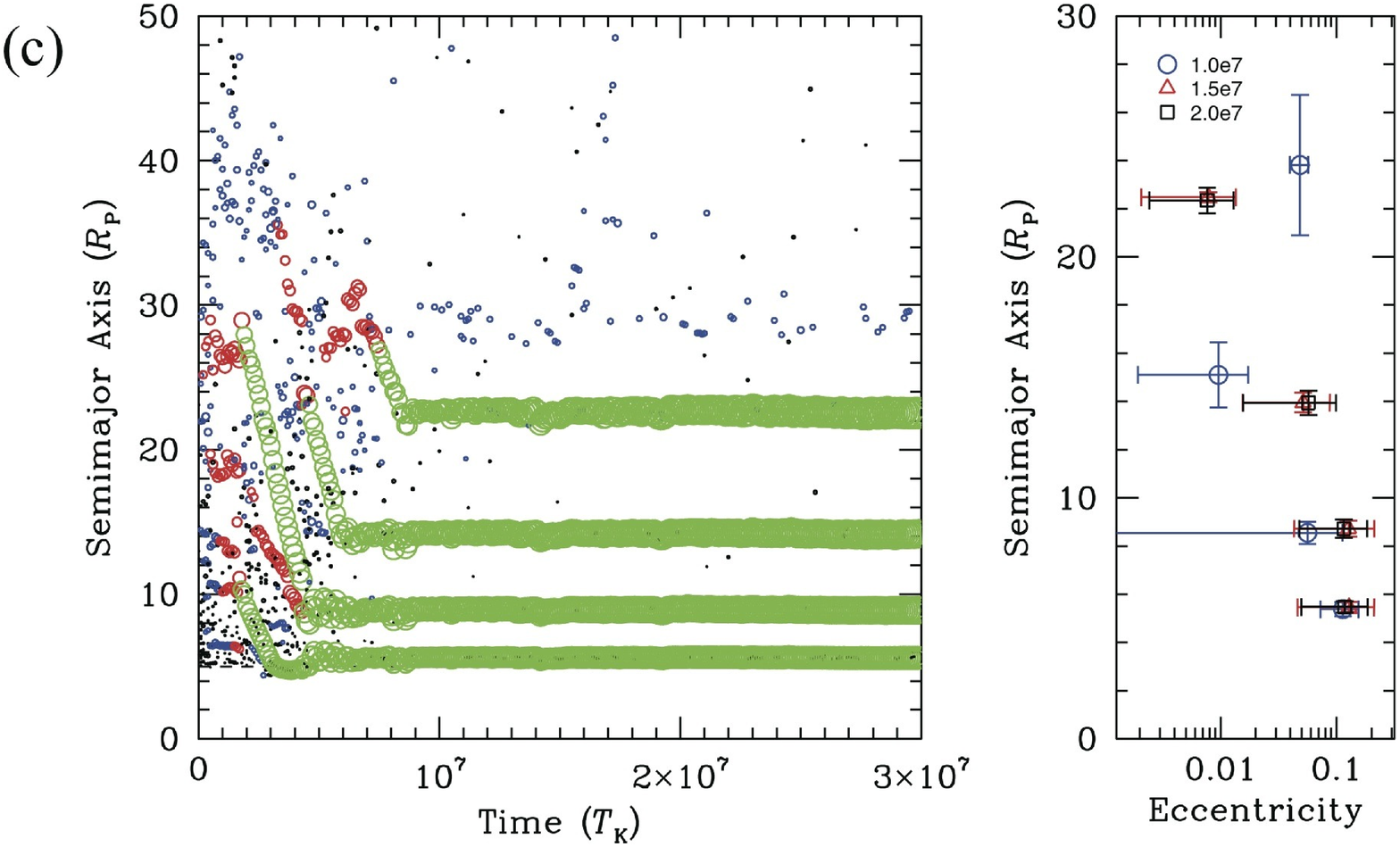}
\plotone{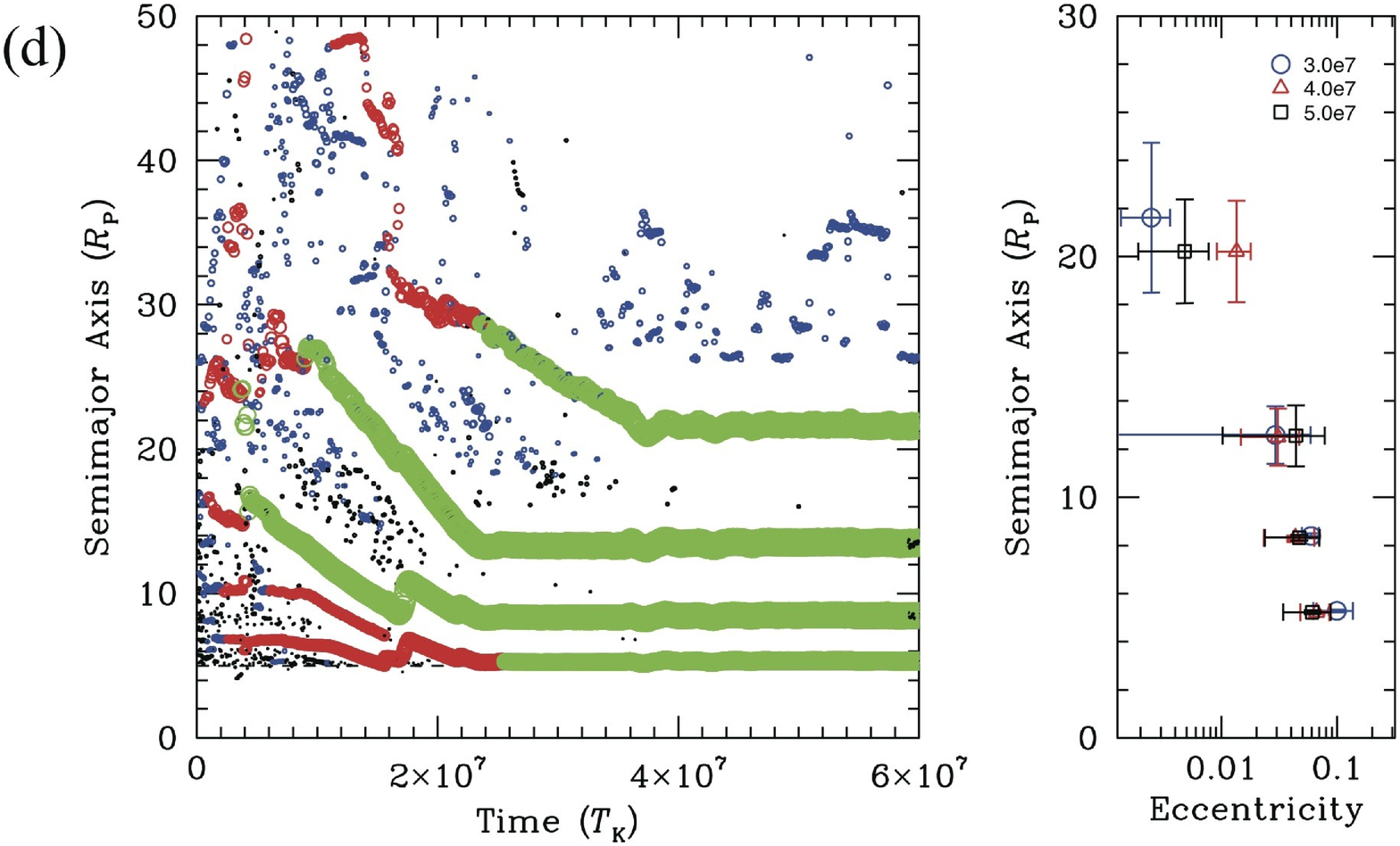}
\plotone{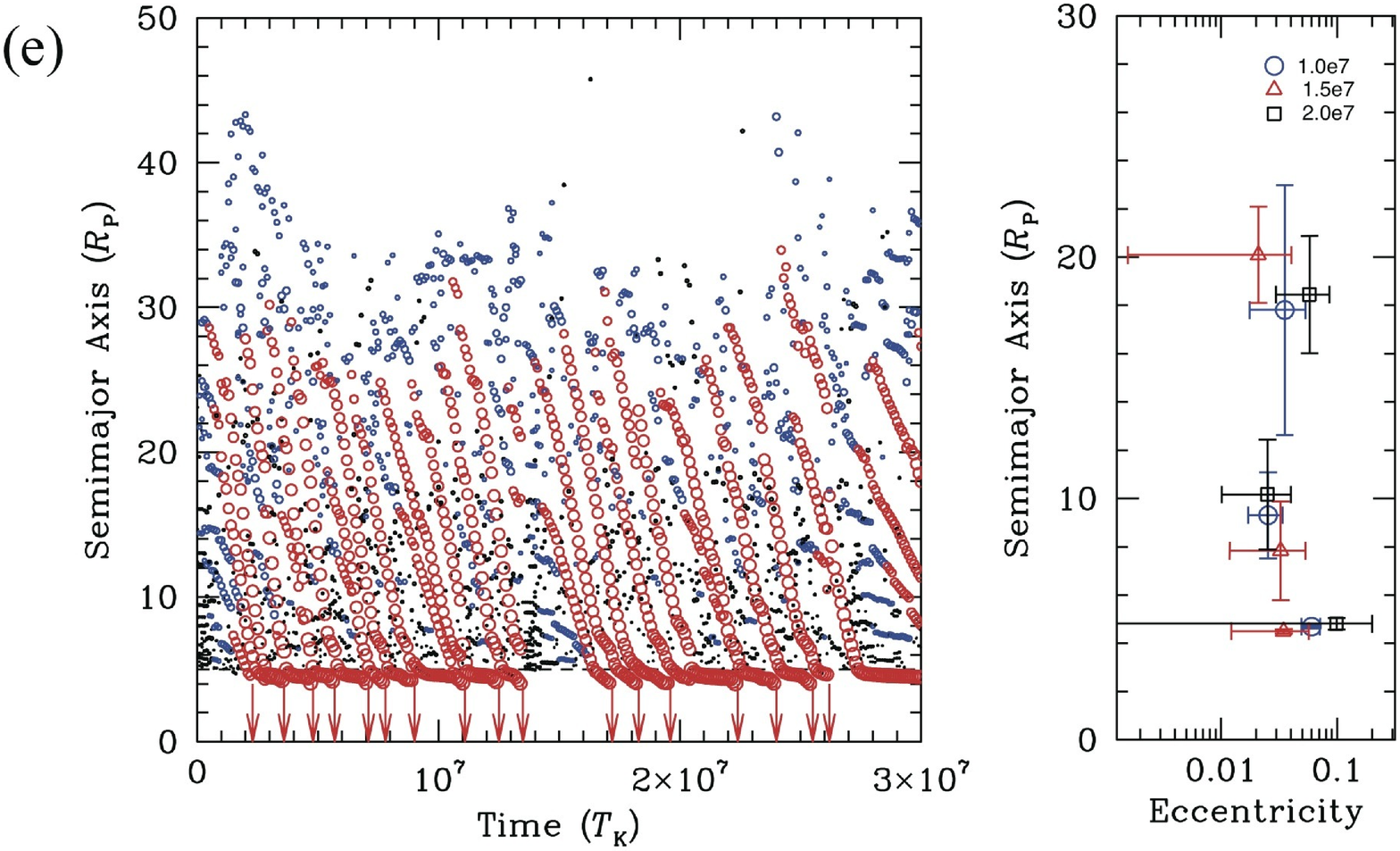}
\caption{Results of \textit{N}-body simulations of (a) model~3, (b) model~4,
(c) model~5, (d) model~6, and (e) model~7. Left: Same as that described in 
Figure~\ref{fig:t_a_run1},
with orbital evolution for representative runs. Right: Same as that described in 
Figure~\ref{fig:e_a},
with averaged orbital distributions.
}
\label{fig:results_parameters}
\end{figure}

We present the results of \textit{N}-body simulations with various parameters
in Figures~\ref{fig:results_parameters}(a)-(e). The model parameters for each
simulation are summarized in Table~\ref{tbl:parameters}. The left panels of
Figures~\ref{fig:results_parameters} show orbital evolution of each representative
run, while the right panels display averaged orbital distributions of the four innermost
satellites at each time over three runs. An exception is Figure~\ref{fig:results_parameters}(e)
in which only the three innermost satellites are plotted.
The times are selected after the time at which the innermost satellite
reaches the disk inner edge.

Figure~\ref{fig:results_parameters}(a) shows the case of model~3, in which the gas surface 
density is 10 times higher than that adopted in model~1. Although the damping
timescales of the eccentricity and the semimajor axis due to the gravitational drag
become 10 times shorter, the trend agrees with that of model~1.
Three to five satellites are lined up outside the disk inner edge, and they
generally exhibit 2:1 mean motion commensurabilities.
The critical mass for starting migration drops to $\simeq 2 \times 10^{-5}~M_{\rm P}$,
and the masses of satellites trapped near the edge are comparable to
those of the Galilean satellites ($\simeq 5 \times 10^{-5}~M_{\rm P}$)
before $t = 6 \times 10^6~T_{\rm K}$. After this time, the masses continue to increase
as the satellites accrete inflowing solid materials. At
$t= 3 \times 10^7~T_{\rm K}$, the mass of the satellites that are
trapped by the edge become $\simeq 3 \times 10^{-4}~M_{\rm P}$,
which is consistent with model~1 because
the solid inflow rate is the same as that of model~1 ($f_{\rm d,in}= 1$).
Despite the increase in
the eccentricity damping force, the eccentricities of the satellites captured into
resonances are nearly comparable to those in model~1, because
the eccentricities of the satellites exhibiting the eccentricity trap are
given as a function of $t_e/t_a$ \citep{ogihara_etal10}. 

Figures~\ref{fig:results_parameters}(b) and (c) show the results of models 4 and 5,
respectively, in which the inflow rate of solid materials is changed. The critical mass for
migration is proportional to $f_{\rm d,in}^{3/5}$; thus, the change of the inflow rate
by a factor of 2 leads to a variation of the critical mass by a factor of $2^{3/5}$.
The masses of satellites that are trapped by the edge at $3 \times 10^7~T_{\rm K}$
are $\simeq 1.5 \times 10^{-4}~M_{\rm P}$ (model~4) and 
$\simeq 6 \times 10^{-4}~M_{\rm P}$ (model~5).
Although the masses of the migrating satellites change, altering
the migration rate, the overall results are hardly affected. In particular, three
to four satellites are captured into the 2:1 mean-motion resonance as is found
in the fiducial model. 

Figures~\ref{fig:results_parameters}(d) and (e) show the results of models 6 and 7,
in which the type I migration speed is reduced (model~6) and increased
(model~7) from model~1 by a factor of 10. 
The critical mass is changes; however, the orbital configuration in
model~6, in which approximately four satellites in the 2:1 resonance are trapped by the edge, 
is approximately the same as in model~1. The masses of satellites trapped by the edge 
at $3 \times 10^7~T_{\rm K}$ are also comparable to those in model~1.
However, the eccentricity trap is 
no longer effective in model~7, which results in orbital distributions of 
satellites that differ significantly from those in other models.

\subsection{Resonances and Number of Satellites}
\label{sec:number}
\begin{figure}[htbp]
\epsscale{1.2}
\plotone{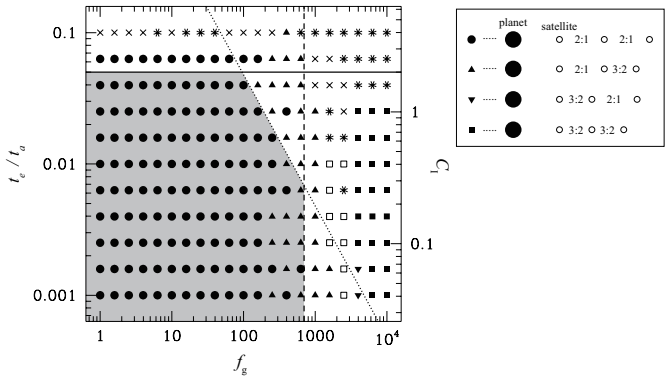}
\caption{Results of three body calculations for various values of $f_{\rm g}$
and $t_e/t_a$. Filled circles indicate that all three satellites are captured 
into the 2:1 resonance outside the disk inner edge. Solid, dashed, and dotted
lines represent semianalytically estimated limits by Equations~(\ref{eq:e_trap2}), 
(\ref{eq:trap_e}), and (\ref{eq:trap_a}), respectively.
}
\label{fig:small_body}
\end{figure}

In this subsection,
we discuss properties of remaining satellites in detail.
To further discuss the resonant configuration and the
total number of formed satellites over a wide parameter range,
we calculate the orbital evolution of hypothetical three satellite systems.
Here we show that capture into 2:1 resonance is robust for a reasonable
range of parameters.
The masses of satellites are set to $10^{-4}~M_{\rm P}$.
The gas surface density $f_{\rm g}$, which affects damping rates
of $e$ and $a$, and the migration speed $C_{\rm I}$ (hence $t_e/t_a$)
are treated as free parameters. The results are summarized in 
Figure~\ref{fig:small_body}.
Filled circles and filled squares in the figure indicate that all the three
satellites exhibit the eccentricity trap and are captured into 
the 2:1 and 3:2 resonances, respectively.
Filled triangles represent the cases in which the satellites 
exhibit the eccentricity trap but the inner pairs are in the 2:1 resonance
and the outer pairs are in the 3:2 resonances. Inverted 
triangles represent inner pairs in the 3:2 resonance and
outer pairs in the 2:1 resonances.
Open squares represent transitional cases from the filled triangles
to the filled squares in which inner and outer
pairs are temporarily captured into the 2:1 and 3:2 resonances, respectively, outside
the disk edge; after a while, configurations of the inner pairs are
converted to the 3:2 resonance when the innermost satellites
move inside the edge.
Crosses represent cases in which the eccentricity trap
is not realized, although the satellites are captured into the 2:1 or 3:2 resonances.
Asterisks show cases of satellites collisions. 

We find that the region (filled circles) in which satellites are captured into the 2:1
resonance outside the disk edge is vast.
In fact, even with rapid migration ($C_{\rm I}\simeq2.5$) and a high gas
surface density ($f_{\rm g} \simeq 10^2$; $\Sigma_{\rm g}\simeq 10^4~{\rm g~cm^{-2}}$),
final configurations are identical.
This region represents the necessary condition for the formation of the Galilean
satellites, suggesting that a Galilean-like configuration in which the three satellites are
in the 2:1 mean motion resonance can be formed with near certainty if 
the disk inner edge is sufficiently sharp.
In some runs, we observe that the Laplace angle $\theta_5$ librates about
some fixed values.

It is difficult to derive analytical formulae that represent
the filled circle region because resonant capture for equal-mass
multiple systems has not been analytically discussed.
Therefore, we obtain semianalytical conditions
by using simplified treatments.
From Figure~\ref{fig:small_body}, we find that the following three
conditions are necessary for capture into the 
2:1 resonance outside the disk edge: (i)~satellites should be trapped towing to 
the eccentricity trap to avoid falling onto the planet, (ii)~eccentricity damping
should not be excessively strong, and (iii)~the migration speed should not be
extremely rapid. Next we derive semianalytical formulae for each condition.

Initially, the condition (i) is considered.
For the eccentricity trap to occur, a constraint on $t_e/t_a$
(hence $C_{\rm I}$) is derived by an argument similar to that provided in 
\citet{ogihara_etal10}. If the edge torque, which is exerted on the 
innermost satellite because of the eccentricity damping force, balances
migration torques of the outer satellites, then the satellites can be trapped
by the edge. The trapping condition is obtained from Equation~(32)
of \citet{ogihara_etal10}:
\begin{equation}
\frac{-A^{c}_{\theta}}{0.78}\frac{e}{\pi} \frac{1}{t_e}-
\frac{1}{2t_a} -\frac{n-1}{2t_a}>0,
\label{eq:e_trap1}
\end{equation}
where $A^{c}_{\theta}(= -0.868)$ is the numerical coefficient \citep{tanaka04}, 
$e$ is the eccentricity of the innermost satellite, and $n$ is the 
number of trapped satellites. 
The first term is the edge torque,
the second term is the migration torque on the innermost satellite,
and the third term is the resonant torque from outer satellites.
Here, we assume that the satellites have nearly equal masses for
simplicity.
(Note that we used a more accurate expression for the edge torque than that
in Equation~(32) of \citet{ogihara_etal10}.)

From numerical calculations, eccentricities excited by resonances
are well fitted by
\begin{equation}
e\simeq 0.6 \left(\frac{t_e}{t_a}\right)^{1/2} \left(\frac{n}{3}\right)^{1/2}.
\label{eq:e}
\end{equation}
Substituting Equation~(\ref{eq:e}) into Equation~(\ref{eq:e_trap1}),
the following constraint on $t_e/t_a$ is obtained:
\begin{equation}
\frac{t_e}{t_a}<0.05  \left(\frac{n}{3}\right) \left(\frac{2}{n-1}\right)^{2},
\label{eq:e_trap2}
\end{equation}
where the second term of Equation~(\ref{eq:e_trap1}) is ignored.
The orbit of the innermost satellite partially enters the inner cavity,
in which the satellite is not affected by the type I torque. In this case, 
the magnitude of type I migration torque on the innermost satellite
should be smaller than $1/2t_a$, and the third term dominates the second 
term \citep{ogihara_etal10}.
The capture constraint by the eccentricity trap is plotted with a solid line
in Figure~\ref{fig:small_body}. This condition is consistent with the
results of orbital calculations. 

Then, from condition (ii), 
we derive a constraint by comparing timescales of excitation and
eccentricity damping. The eccentricity excitation by a single
distant encounter of two satellites on approximately circular orbits is given by
\citep{hasegawa_nakazawa90}
\begin{equation}
\delta e \simeq 6.7 \left(\frac{b}{r_{\rm H}}\right)^{-2}
\left(\frac{r}{r_{\rm H}}\right)^{-1},
\end{equation}
where $b$ is the difference in the semimajor axes of the two satellites.
Because the distant encounter occurs at every synodic period
\begin{equation}
T_{\rm syn} \simeq \frac{2\pi r}{3b\Omega /2},
\end{equation}
the variation of eccentricity per unit time is expressed as
\begin{equation}
\left.\frac{de}{dt}\right|_{\rm scat} \simeq \frac{\delta e}{T_{\rm syn}}\simeq
10\left(\frac{b}{r_{\rm H}}\right)^{-1}
\left(\frac{r}{r_{\rm H}}\right)^{-2} \frac{\Omega}{2\pi}.
\label{eq:de_dt_scat}
\end{equation}
For first-order resonances of $j:j-1$, the orbital separation is 
expressed as
\begin{equation}
b = \left[
\left(\frac{j}{j-1}\right)^{2/3}-1 
\right]a,
\end{equation}
then, Equation~(\ref{eq:de_dt_scat}) is reduced to
\begin{eqnarray}
\left.\frac{de}{dt}\right|_{\rm scat} &\simeq& 1.7 \times 10^{-2} 
\left(\frac{M}{10^{-4}~M_{\rm P}}\right)
\left(\frac{r}{20 R_{\rm P}}\right)^{-3/2}
\nonumber \\ && \times
\left(\frac{R_{\rm P}}{R_{\rm J}}\right)^{-3/2}~{\rm yr}^{-1},
\label{eq:de_dt_scat2}
\end{eqnarray}
where $j= 2$ is assumed in the final transformation.
A necessary condition for capture into a resonance is that
$e$-excitation
due to distant encounters should be larger than $e$-damping
due to the gas drag described in Equation~(\ref{eq:Tanaka_e_s}):
\begin{equation}
\left. \frac{de}{dt}\right|_{\rm scat} > \frac{e}{t_e}.
\end{equation}
This leads to the following constraint on $f_{\rm g}$:
\begin{eqnarray}
f_{\rm g} &<& \frac{70}{e} \left(\frac{M_{\rm P}}{M_{\rm J}}\right)^{-3/2}
\left(\frac{r}{20 R_{\rm P}}\right)^{-3/4}
\left(\frac{R_{\rm P}}{R_{\rm J}}\right)^{-3/4}
\nonumber \\ && \times
\left(\frac{\tau_{\rm G}}{5\times 10^6~{\rm yr}}\right)^{-1/2}.
\label{eq:trap_e}
\end{eqnarray}
From Figure~\ref{fig:small_body}, we find that the capture condition
for the 2:1 resonance is well explained by $f_{\rm g} < 700$,
in which $e\simeq 0.1$ is used. This capture constraint is plotted
with a dashed line in Figure~\ref{fig:small_body}.

Finally, from condition (iii), 
a necessary constraint condition on the migration speed
is derived. According to both analytical \citep{friedland01} and numerical
\citep{ida_etal00,wyatt03} studies of resonant trapping, capture
probability depends on the mass and the semimajor axis. 
The dependence of the critical migration rate, which can be represented by $C_{\rm I,crit}$,
for trapping is given as $C_{\rm I,crit} \propto (M/M_{\rm P})^{-4/3} \Omega^{-1}$.
Then, we adjust the coefficient by our numerical calculations.
The numerical result shown in Figure~\ref{fig:small_body} is fitted as
\begin{equation}
C_{\rm I}<200 f_{\rm g}^{-1} \left(\frac{M}{10^{-4}~M_{\rm P}}\right)^{-4/3},
\label{eq:trap_a}
\end{equation}
which is the critical migration timescale of the three satellites for the 2:1 resonance, 
and it is plotted with a dotted line in Figure~\ref{fig:small_body}.

\begin{figure}[htbp]
\epsscale{1.2}
\plotone{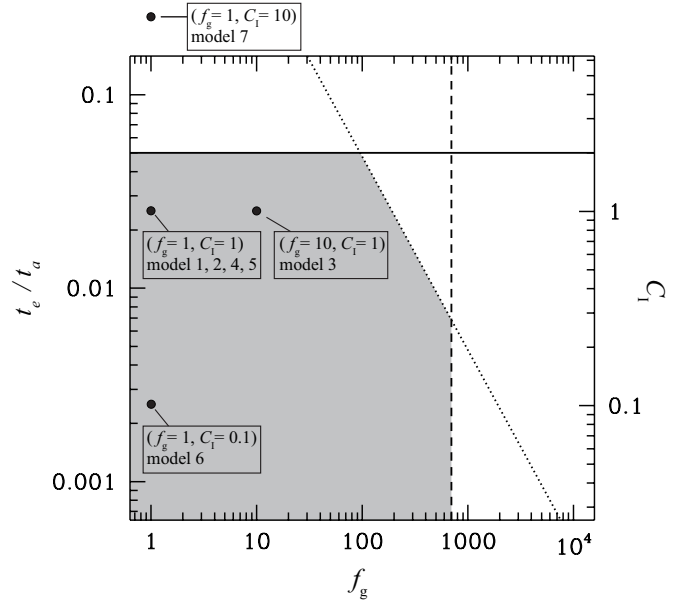}
\caption{Model parameters adopted in \textit{N}-body simulations.
The parameters of model~7 are outside the shaded region, which are derived by
a semianalytical argument.
}
\label{fig:model}
\end{figure}

The enclosed shaded region in Figure~\ref{fig:small_body}
represents the condition for formation of Galilean-like satellites.
The results of three satellite calculations and the semi-analytical arguments
are consistent with the results of \textit{N}-body simulations. 
Figure~\ref{fig:model} indicates the parameters used in our \textit{N}-body
simulations on the gas surface density ($f_{\rm g}$) and the migration speed
($C_{\rm I}$) plane. The shaded region corresponds to that
in Figure~\ref{fig:small_body}.
In case of models~1-6, the parameters are in a good region for reproducing the
number and the resonant relationships of the Galilean satellites, while 
the parameter values of model~7 is far from the eccentricity trapping limit.

We also compare our results with previous studies, in which the
Hamiltonian model is used in case of rapid migration (\citealt{quillen06}; \citealt{mustill10}).
These studies have considered the circular restricted three-body problem with a massive
planet and a massless test particle orbiting a central star to 
derive capture probability for first- and second-order resonances.
Because the bodies with comparable masses are captured into resonances
in our simulation and the effect of eccentricity damping is included,
it is not possible to directly apply their predictions to our case.
However, we find that their results are approximately consistent with
our \textit{N}-body results and hence our semianalytical arguments.
In \citet{mustill10}, capture probability is derived as a function of the
generalized momentum $J$ and the migration rate $\dot{\beta}$.
Since $J \sim 10^{-1}$ and $\dot{\beta}\sim 10^{-2}$
are derived in the case of model~1, it is determined from Figures~2 and 11 
in \citet{mustill10} that capture into the 2:1 resonance is highly likely. 
Therefore, our findings of a robust 2:1 resonance formation are also 
confirmed by the Hamiltonian model.
In addition, \citet{mustill10} documented the following trends, which is also observed in our 
calculations:
capture probability decreases with a decrease in mass.
If the migration rate increases by $10^3$ times of that in model~1, 
bodies tend to be captured into the closer 3:2 resonance.
These features therefore agree with our semianalytical estimates.
Further investigation of the resonant capture for comparable masses
including eccentricity damping is expected in future work.

Finally, we note that since the satellites are captured into the 2:1 resonance,
the orbital difference $b$ is generally 
$\simeq 10-15~r_{\rm H}$, which is larger than $b = 5~r_{\rm H}$ 
assumed in \citet{sasaki_etal10}. In their calculations, small satellites are trapped 
between large satellites; as a result, orbital differences between large 
satellites become $\sim 15~r_{\rm H}$. Their assumption is valid 
for relatively fast migration cases (e.g., planet formation); therefore
slight improvement is needed. This is also discussed in Section~\ref{sec:saturn}.

\subsection{Mass}
\label{sec:mass}
\begin{figure}[htbp]
\epsscale{1.2}
\plotone{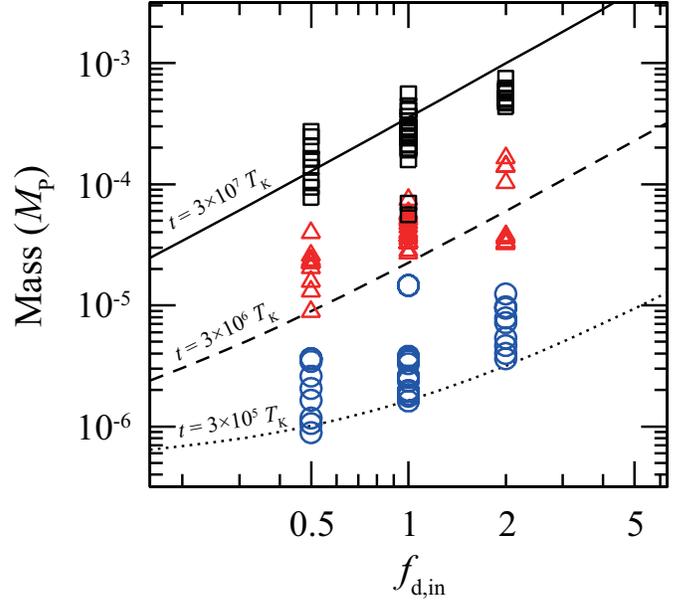}
\caption{Mass of the largest three satellites at $3\times 10^5~T_{\rm K}$
(blue circles), $3\times 10^6~T_{\rm K}$ (red triangles), and 
$3\times 10^7~T_{\rm K}$ (black squares) for models~1 and 3 ($f_{\rm g}= 1$),
model~4 ($f_{\rm g}= 0.5$), and model~5 ($f_{\rm g}= 2$). The dotted,
dashed, and solid lines are estimates of Equation~(\ref{eq:m_evolution}) at
$3\times 10^5~T_{\rm K}$, $3\times 10^6~T_{\rm K}$, and $3\times 10^7~T_{\rm K}$,
respectively.
}
\label{fig:results_parameters2}
\end{figure}

In this subsection, we discuss the satellite mass by using semianalytical 
estimates. As mentioned in Section~\ref{sec:growth}, satellites
grow according to the accretion timescale of Equation~(\ref{eq:acc}). 
Adding the exponential decay of the gas inflow, we can obtain the 
satellite mass as a function of time by integrating Equation~(\ref{eq:acc}):
\begin{eqnarray}
\frac{M(t)}{10^{-4}~M_{\rm P}}&=&\biggl[
3.7 f_{\rm d,in} \eta_{\rm ice} \left(\frac{r}{20~R_{\rm P}}\right)^2
\left(\frac{\tau_{\rm dep}}{10^6~{\rm yrs}}\right)
\nonumber \\ && \times
\left\{1-\exp\left(\frac{-t}{\tau_{\rm dep}}\right)\right\}
\nonumber \\ && \times
+ \left(\frac{M_0}{10^{-4}~M_{\rm P}}\right)^{2/3}
\biggr]^{3/2},
\label{eq:m_evolution}
\end{eqnarray}
where $M_0$ is the initial mass. The time dependence of $r$ is neglected.
Figure~\ref{fig:results_parameters2}
plots the mass of the largest three satellites as a function of $f_{\rm d,in}$
for 12 runs (models~1, 3-5). The circles,
triangles, and squares express the mass at $3\times10^5$,
$3\times10^6$, and $3\times10^7~T_{\rm K}$, respectively.
The estimates of Equation~(\ref{eq:m_evolution}) are also drawn
with the dotted line ($3\times10^5~T_{\rm K}$), 
the dashed line ($3\times10^6~T_{\rm K}$), and the solid line
($3\times10^7~T_{\rm K}$), where $r=20~R_{\rm P}$ is substituted.
Because satellites migrate inward from
the original locations, we neglect $r$-dependence.
We find that the results are reasonably consistent with the estimates.
However, it should be noted that although $\Delta r = 10~r_{\rm H}$ is 
assumed in deriving Equation~(\ref{eq:acc}), the orbital separation
of satellites trapped by the edge can decrease to
$\lesssim 10~r_{\rm H}$ with an increase in mass, and $\Delta r$ becomes
independent of $M$. Thus, the power-law index in the right-hand-side of
Equation~(\ref{eq:m_evolution}) decreases to unity. In fact, 
the slope at $3\times10^7~T_{\rm K}$ is close to the linear dependence 
of $f_{\rm d,in}$.

\citet{canup06} showed that the mass of satellites and the total mass
in a system are both regulated by the balance between the supply of the 
inflowing materials and orbital decay due to type I migration. The critical mass
is derived in Equation~(\ref{eq:m_crit}). However, in our case in which
the inner cavity exists, the total mass regulation
does not work effectively unless the eccentricity trap is inhibited. 
For a case in which the number of trapped satellites by the edge
is equal to the critical number for the eccentricity trap, the configuration
is destroyed and some satellites are lost to the planet if an additional
satellite migrates and is trapped in a resonance.
In other cases, the total mass continues to increase.
As discussed in \citet{sasaki_etal10}, when the total mass 
of the trapped satellites exceeds the disk mass, the satellites may 
be released to the host planet, and mass regulation occurs.
In addition, recent magnetohydrodynamic (MHD) simulations (e.g., \citealt{romanova_etal08}) 
suggest that the inner cavity may be buried by intermittent gas accretion onto the planet,
and satellites trapped by the edge would episodically fall.
These effects should be examined in detail through MHD simulations. 

With an analytical estimate of the growth timescale, model parameters
can be constrained for the formation of the Galilean satellites.
To prevent these parameters from exceeding the masses of the Galilean satellites, 
$M_{\rm crit} \leq 5 \times 10^{-5}~M_{\rm P}$ should be satisfied
as a necessary condition, which leads to (Equation~(\ref{eq:m_crit})):
\begin{eqnarray}
\frac{\eta_{\rm ice}f_{\rm d,in}}{C_{\rm I}f_{\rm g}} &\leq& 0.5
\left(\frac{M_{\rm P}}{M_{\rm J}}\right)
\left(\frac{r}{20~R_{\rm P}}\right)^{-5/2}
\left(\frac{R_{\rm P}}{R_{\rm J}}\right)^{-1/2}
\nonumber \\ && \times
\left(\frac{\tau_{\rm G}}{5 \times 10^6~{\rm yr}}\right)^{1/4},
\end{eqnarray}
where we neglect the impact erosion of ice.
In addition, to be consistent with Callisto's partially differentiated state,
an accretion time longer than $\sim 5 \times 10^5~{\rm years}$ is implied to enable
$\eta_{\rm ice}f_{\rm d,in}<1$ at the time of Callisto's accretion
(Equation~(\ref{eq:acc})).

\section{FORMATION OF SATURNIAN SYSTEM}
\label{sec:saturn}
\citet{sasaki_etal10} proposed that Jupiter opened a gap in the 
protoplanetary disk to halt its growth, and Saturn did not. The
circum-Saturnian disk gradually dissipated on the global depletion timescale of
the protoplanetary disk; therefore, in the final stage of satellite accretion around 
Saturn, the circum-Saturnian disk did not have an inner cavity because a low
accretion rate would lead to a weak magnetic field, as discussed in Section~\ref{sec:model}. 
This condition
is comparable with that adopted in \citet{canup06}; however, properties of the
Saturnian system are not commonly reproduced either by their \textit{N}-body
simulations or our preliminary \textit{N}-body calculations without the inner cavity. That is,
the average number of final satellites is $\sim 3-4$, while the number of
large satellites ($\geq 10^{-5}~M_{\rm P}$) in the current
Saturnian system is one (Titan). 
Hence, the solid material distribution produced by \textit{N}-body calculations
is also not consistent in the current Saturnian system.

According to Equation~(\ref{eq:acc}), the growth timescale of satellites decreases
with distance from the central planet; therefore, satellite seeds
predominantly develop in the outer region ($\gtrsim 20~R_{\rm P}$), 
and the satellites exhibit inward migration.
If the migration rate significantly increases during migration, the migrating satellites
rapidly fall onto the planet, and as a result, a single satellite always exists in
Titan's orbit. 
\citet{sasaki_etal10} assumed accelerated migration and showed that 
one or two satellites remain in the end; however, their method for generating 
satellite seeds is not appropriate for satellite formation.
Because the migration timescale in protosatellite disks is proportional to $r^{1/2}$
(Equation~(\ref{eq:Tanaka_a_s})), the migration rate is not accelerated
($da/dt = a/t_a \simeq a^{1/2}$) during migration, which is observed in 
\textit{N}-body simulations of \citet{canup06} and in this study.
It should be noted that planets in a minimum-mass nebula, in which
a steeper surface density profile ($\Sigma_{\rm g} \propto r^{3/2}$) is 
considered, undergo accelerated migration, because $t_a \propto r^{3/2}$.
Thus, we suggest in this section several possibilities to reconcile the inconsistency.

In our calculations and also in those of \citet{canup06} and \citet{sasaki_etal10},
the radial dependence of the gas inflow is ignored, which results in a 
relatively gentle gradient of gas surface density ($\Sigma_{\rm g}
\propto r^{-3/4}$). If the gas inflow flux is inversely proportional to
$r$ ($F_{\rm in} \propto r^{-\gamma}; \gamma>0$), the gradient of gas
surface density becomes steep; recent hydrodynamic
simulations by \citet{tanigawa_etal11} suggest that $\gamma \sim 1$.
In addition, photoevaporation can produce the steep gradient of gas density
\citep{mitchell_stewart11}.
The type I migration speed increases
in the region close to the central planet so that the satellites that begin
to undergo migration are rapidly lost to the planet. Our preliminary
\textit{N}-body simulations that include this effect suggest that the number
of satellites decreases with an increase in $\gamma$. More detailed 
calculations are required, and it is also important to examine the radial
and time dependence of the inflow flux by a high-resolution hydrodynamic 
simulation. 

Another possibility is that the solid density is locally increased around
Titan's orbit ($\simeq 20~R_{\rm P}$), leading to the preferential growth
of satellites in the region.
The radial distribution of solid inflow and satellitesimal formation should 
be investigated. A detailed discussion of the Saturnian system formation
will be presented in a separate paper.

\section{CONCLUSIONS}
\label{sec:conclusion}
We have investigated the formation of regular satellites around giant planets
using direct \textit{N}-body simulations, including the effects of the eccentricity trap
near the disk edge and the damping of the semimajor axis and eccentricity.
Through these simulations, we confirmed the following scenarios of the
final stage of a satellite formation in a steady accretion disk with an inner cavity
and a gas/solid inflow from a circumstellar disk:
\begin{enumerate}
\item Within the inflow region, protosatellites accrete from solid materials in their 
feeding zones. When the satellites grow to the critical mass for migration, 
they begin to exhibit  orbital decay. Because the width of the feeding zone increases with
an increase in distance from the central planet and disk surface density has
a relatively weak radial dependence ($\Sigma_{\rm g} \propto r^{-3/4}$), 
satellites in outer orbits grow larger than inner satellites.
\item When the satellites reach the inner disk edge, they cease inward migration.
The subsequently migrating satellites are captured into mean motion resonances 
with the inner satellites. If the edge torque exerted on the innermost satellite balances
the migration torque, the satellites are trapped near the edge,
which is called as the eccentricity trap. 
\item Satellites with the critical mass migrate inward to be trapped in successive
resonances. If the number of trapped satellites exceeds the
critical number for the eccentricity trap, the orbital configuration of the resonant 
convoy is disrupted, resulting in a loss of the inner satellites to the planet.
That is, the total number of trapped satellites is self-regulated.
\item The disk gas of circumplanetary protosatellite disks quickly ($\sim 10^3~{\rm years}$)
vanishes because of truncation of inflowing gas by the gap opening along the planet's orbit
in the circumstellar disk, and resonant configuration is frozen at that time. 
Although the eccentricities of trapped satellites are excited by resonances,
they are generally damped to $\lesssim 0.01$ through long-term tidal dissipation 
within the satellites.
\end{enumerate}

We tracked the compositional evolution of satellites in \textit{N}-body simulations
including the ice line beyond which water is condensed. As migration of icy 
satellites from the outer region ceases because of the eccentricity trap, 
inner satellites avoid contamination by water-rich materials, resulting in
a radial compositional gradient. This composition is described as an 
increase in icy components with distance. 
In addition, it is suggested that to reproduce the 
compositional gradient, the gap opening needs to occur while the satellites
are trapped by the edge via the eccentricity trap.
An additional pathway for producing the compositional gradient
is also suggested such that the volatile materials of inner satellites would be 
evaporated because of high-speed collisions during their accretion.

Furthermore, we examine the relationship between the characteristics of the final satellites 
and the model parameters by numerical simulations and semianalytical 
arguments to evaluate the probability of the Galilean-like
satellite formation. The physical properties of the Galilean moons provide the following
constraints on the model parameters. The number of satellites that can be
trapped by the eccentricity trap restricts the ratio of the $e$-damping timescale
versus the $a$-damping timescale:
\begin{equation}
\frac{t_e}{t_a}<0.05,
\label{eq:cond_etrap}
\end{equation}
where $n= 3$ is substituted into Equation~(\ref{eq:e_trap2}).
Equation~(\ref{eq:cond_etrap}) is given as $(c_s/v_{\rm K})^2 \lesssim 0.01/C_{\rm I}$,
which is usually satisfied in the protosatellite disk with the temperature of
Equation~(\ref{eq:temp}).
The number of final satellites also depends on the inflow region of 
the solid materials, although the actual inflow region of gas and dust 
for the circum-Jovian orbit is probably 
consistent with our adopted theory \citep{machida09}.
By considering capture into resonances, constraints on the $e$-damping and 
$a$-damping timescales are derived. A necessary condition for capture into the 2:1 
resonance is
\begin{eqnarray}
f_{\rm g} <  700.
\end{eqnarray}
This parameter can be easily satisfied in the gas-starved disk ($f_{\rm g}= 1$), 
and this condition implies that an actively-supplied disk with either a much 
lower viscosity and/or a more rapid inflow rate than considered in previous works 
can also produce 2:1 resonant orbits.
An additional condition for satellites with masses of $5 \times 10^{-5}~M_{\rm P}$ is
\begin{equation}
C_{\rm I}<500 f_{\rm g}^{-1}.
\end{equation}
This parameter can also be satisfied because $C_{\rm I} \leq 1$.
We therefore find that formation of Galilean-like satellites is robust, because
even with fast migration and a high gas surface density over three satellites,
they are captured into the 2:1 mean motion resonance outside the disk inner edge.
The parameters for growth and migration are also constrained by considering
that the satellite mass does not exceed the critical mass for migration:
\begin{equation}
\frac{\eta_{\rm ice} f_{\rm d,in}}{C_{\rm I} f_{\rm g}}\leq 0.5.
\end{equation}
It is also inferred that the solid inflow is decreased ($\eta _{\rm ice} f_{\rm d,in}<1$)
at the time of Callisto's formation. In addition, to prevent Callisto from being captured
into mean motion resonances, the gap opening timing of Jupiter should be comparable 
to the timing of the completion of Callisto formation. This theory is consistent not only with the gap
opening time, which was inferred from the discussion of composition, but also with the
time assumed in our calculation.

We finally provide theoretical predictions for characteristics of exomoons,
which should be detected in the near future. 
If the disk inner edge width is sufficiently sharp, the eccentricity
trap is likely to occur because the value of $t_e/t_a$ is generally small 
even in the case that applies the full type I migration rate from the linear 
theory. The number of trapped satellites by the disk edge 
would be less than $\sim 10$ because the inflow region of solid material 
is probably consistent with that in the adopted theory \citep{machida09}.
In the last stage of satellite accretion, the gas surface density would
presumably be less than $10^5~{\rm g~cm^{-2}}$ so that the 
damping timescales of eccentricity and semimajor axis are sufficiently long
to enable the capture of satellites into the 2:1 mean-motion resonance,
unless the mass of the satellite is significantly small.

These properties of multiple satellite formation and capture into resonances
imply that the satellites can be substantially affected by tidal heating.
The satellites that have a sufficient mass to retain water in the HZ 
are potentially habitable. Furthermore, according to tidal heating,
their host planet is not required to be in the HZ; thus, the probability of 
forming Europa-like habitable satellites, which have tidally heated oceans,
would be high. From the perspective of tidally heated habitable moons, it is probable that 
satellites in relatively outer orbits retain water compared to close-in
moons, which are strongly heated. Tidal heating is extremely
important for habitable moons; therefore, further study
of the evolution of satellites including tidal heating is necessary.
By considering the migration efficiency and the accretion rate, 
the mass of normal satellite size would be $\sim 10^{-4}~M_{\rm P}$, which is
similar to that of the Galilean satellites and Titan. It is possible for observable
exomoons ($\gtrsim 0.2~M_{\oplus} \simeq 6 \times 10^{-4}~M_{\rm J}$)
to form around giant planets more massive than Jupiter.
Satellite systems with a large mass ratio of the satellite to the host planet
are more stable for detecting exomoons \citep{kipping_etal12}, and 
when the factor $\eta_{\rm ice} f_{\rm d,in}/C_{\rm I} f_{\rm g}$ is larger than unity,
larger satellites ($\gtrsim 10^{-4}~M_{\rm P}$) are formed.

We find that several satellites are formed locked in resonances under 
the assumption that the host planet opens a gap around its orbit.
\citet{sasaki_etal10} proposed that if the planet does not create a gap,
satellites are unlikely to form in resonances, such as that observed in Saturnian satellites.
Because a gap opening indicates that the planet may undergo type II
migration, it is expected that satellites systems in mean motion
resonances migrate inward to some extent. 
Therefore, planets that reside in closer orbits probably harbor 
Jovian-system-like moons, although
destabilization by tidal torque \citep{barnes_obrien02} or shrinkage of
Hill sphere \citep{namouni10} may also be important.
In contrast, Saturnian-system-like moons orbit more distant planets.

\subsection*{ACKNOWLEDGMENT}
We thank the anonymous referee for useful comments that improved 
and clarified this manuscript.
We also thank Francis Nimmo and Takanori Sasaki for their fruitful 
discussion and valuable suggestions.
Numerical computations were in part conducted on GRAPE system and
the general-purpose PC farm at Center for Computational Astrophysics, 
CfCA, of National Astronomical Observatory of Japan.
This work is supported by Grant-in-Aid for JSPS Fellows (23004841).

\appendix

\section{Actively Supplied Disk Model}
\label{app:disk}
On the basis of \citet{canup02}, the asymptotic disk gas surface density 
for a steady accretion disk is given by
\begin{equation}
\label{eq:sigma_app}
\Sigma_{\rm g} \simeq  \frac{F_{\rm P}}{3\pi\nu}
\left[ 1-\frac{4}{5} \sqrt{\frac{r_{\rm c}}{r_{\rm d}}} -\frac{1}{5}
\left( \frac{r}{r_{\rm c}} \right)^2
\right] \simeq 0.55 \frac{F_{\rm P}}{3\pi\nu},
\end{equation}
where $r_{\rm d}$ is the outer edge of the diffused-out disk.
The second and third terms in the bracket are correction terms
due to outward diffusion. 

The disk is heated by luminosity from the central planet,
viscous dissipation, and energy dissipation associated with 
the difference between the free-fall energies of the incoming gas
and a Keplerian orbit. We assume that heating
is dominated by viscous dissipation; then, the photospheric
temperature of the disk ($T_d$) is determined by a balance between
viscous heating and blackbody radiation from the
photosphere:
\begin{equation}
\sigma_{\rm SB} T_d^4 \simeq \frac{9}{8}\Omega \nu \Sigma_{\rm g}
\simeq \frac{0.55 \times 3}{8 \pi} \Omega^2 F_{\rm P}
\end{equation}
where $\sigma_{\rm SB}$ is the Stephan-Boltzman constant.
When $F_{\rm P} = M_{\rm P}/\tau_{\rm G}$, 
the temperature is reduced to
\begin{equation}
\label{eq:temp_app}
T_d \simeq 160 \left(\frac{M_{\rm P}}{M_{\rm J}}\right)^{1/2}
\left(\frac{\tau_{\rm G}}{5 \times 10^6 {\rm ~yr}}\right)^{-1/4}
\left(\frac{r}{20 R_{\rm P}}\right)^{-3/4}
\left(\frac{R_{\rm P}}{R_{\rm J}}\right)^{-3/4} {\rm K}.
\end{equation}
The midplane temperature $T$ is approximately given by 
$T \simeq (1+3\tau /8)^{1/4} T_d$, where $\tau$ is the optical
depth. Because we are concerned with the disk temperature for 
evolution of the ice line, we assume that $T \simeq T_d$ to
avoid uncertainly in opacity.

Thus, by adopting the alpha model for disk viscosity
($\nu = \alpha c_s H$), the gas surface density of the disk 
becomes
\begin{eqnarray}
\Sigma_{\rm g} \simeq 0.55 \frac{F_{\rm P}}{3\pi\nu}
\simeq 100 f_{\rm g} \left(\frac{M_{\rm P}}{M_{\rm J}}\right)
\left(\frac{r}{20R_{\rm P}}\right)^{-3/4}
\left(\frac{R_{\rm P}}{R_{\rm J}}\right)^{-3/4} {\rm ~g~cm^{-2}},
\end{eqnarray}
\begin{eqnarray}
f_{\rm g}  \equiv \left( \frac{\alpha}{5\times10^{-3}}\right)^{-1}
\left(\frac{\tau_{\rm G}}{5\times10^6 {\rm ~yr}}\right)^{-3/4}.
\end{eqnarray}

\section{Growth Timescale}
\label{app:growth}
The growth rate of satellites is determined by the velocity dispersion 
$\sigma$ and the spatial density of satellitesimals $\rho_{\rm d}$. When velocity
dispersion is smaller than the surface escape velocity of the satellite
core, the accretion rate of the core with the mass $M$ at $r$ is
\begin{eqnarray}
\dot{M}\simeq \pi C R^2 \rho_{\rm d} \left(\frac{2GM}{R \sigma^2}\right)\sigma
\simeq \pi C R^2 \Sigma_{\rm d} \Omega_{\rm K} \left(\frac{2GM}{R \sigma^2}\right)
\left(= 2 \pi C \left(\frac{R}{r}\right) \left(\frac{\Sigma_{\rm d} r^2}{M_{\rm P}}\right) 
\left(\frac{v_{\rm K}}{\sigma}\right)^2 M \Omega_{\rm K}\right),
\end{eqnarray}
where $C$ is a numerical factor of 2-3 \citep{stewart_ida00} and 
$\Sigma_{\rm d}$ is the surface density of satellitesimals in the disk.
Because $R/r$ and $\Sigma_{\rm d}r^2/M_{\rm P}$ are large compared to
the corresponding values for planetary growth at $\sim$ 1 AU around the
Sun, the accretion rate is relatively high.
Thus, the accretion timescale is
\begin{eqnarray}
\tau_{\rm acc,core} &=&\frac{M}{\dot{M}}\\
&\simeq& 3 \times10^3 
\left(\frac{C}{2.5}\right)^{-1}
\left(\frac{\Sigma_{\rm d}}{50{\rm ~g~cm^{-2}}}\right)^{-1}
\left(\frac{\rho}{3{\rm ~g~cm^{-3}}}\right)^{1/3}
\left(\frac{M}{10^{-4}M_{\rm P}}\right)^{-1/3}
\left(\frac{M_{\rm P}}{M_{\rm J}}\right)^{1/6}\nonumber \\
&& \times 
\left(\frac{r}{20R_{\rm P}}\right)^{1/2}
\left(\frac{R_{\rm P}}{R_{\rm J}}\right)^{1/2}
\left(\frac{e}{0.1}\right)^{2}~{\rm yr},
\label{eq:acc_core}
\end{eqnarray}
where $\rho$ and $e$ are 
the internal density of the core and the RMS eccentricity of 
satellitesimals, respectively.
Through numerical simulations, we find that 
$\Sigma_{\rm d}\sim 50~{\rm g~cm^{-2}}$ and $e\sim0.1$.

However, in the case in which the core accretion timescale $\tau_{\rm acc,core}$ is 
smaller than the mass inflow timescale $\tau_{\rm acc,inflow}$,
the growth timescale is controlled by the inflow flux of solid materials,
as determined by \citet{canup06}.
Satellites accrete materials across an annulus of width $\Delta r$
and grow in the timescale as
\begin{eqnarray}
\tau_{\rm acc,inflow} &=&\frac{M}{\dot{M}}
\simeq 
\frac{fM}{F_{\rm in} 2\pi r \Delta r}\\
&\simeq& 1.8 \times 10^5 \eta_{\rm ice}^{-1}
f_{\rm d,in}^{-1} 
\left(\frac{r}{20R_{\rm P}}\right)^{-2}
\left(\frac{M}{10^{-4}M_{\rm P}}\right)^{2/3}~{\rm yr},
\label{eq:acc_inflow}
\end{eqnarray}
where $f (= 100/\eta_{\rm ice} f_{\rm d,in})$ is the gas-to-solid mass ratio 
in the inflow, and 
$\Delta r = 10~r_{\rm H}$ is used. 
This derivation differs slightly
from that of \citet{canup06}, in which $\Delta r \sim 2 er$
is assumed. However, the timescale is approximately consistent.
Obviously, $\tau_{\rm acc,inflow}$ is longer than $\tau_{\rm acc,core}$;
thus, $\tau_{\rm acc,inflow}$ can be considered as the growth timescale
of the satellite. 
These estimates are in good agreement with numerical results 
shown in the text.

\section{Resonant Angle}
\label{app:resonant}
The angels are defined as
\begin{eqnarray}
\theta_1 &=& \lambda_1 - 2\lambda_2 + \varpi_1,\\
\theta_2 &=& \lambda_1 - 2\lambda_2 + \varpi_2,\\
\theta_3 &=& \lambda_2 - 2\lambda_3 + \varpi_2,\\
\theta_4 &=& \lambda_2 - 2\lambda_3 + \varpi_3,\\
\theta_5 &=& \theta_2 -\theta_3 =\lambda_1 - 3\lambda_2 + 2\lambda_3,
\end{eqnarray}
where $\lambda_i$ and $\varpi_i$ are mean orbital longitudes and
longitudes of the pericenter, respectively.
The subscripts $i=$ 1, 2, and 3 refer to the innermost satellite (body~1),
the second satellite (body~2), and the third satellite (body~3), respectively. 
These angles represent displacements of longitude of conjunction 
from pericenters; thus, $\theta_1=0^\circ,~\theta_2=180^\circ$
indicates that conjunctions of body~1 and body~2 occur 
when body~1 is near the apocenter and body~2 is near 
the pericenter.
If the angles librate about some fixed values, it is usually 
considered that they are in mean motion resonances.
In addition, the Laplace resonance is characterized by the libration of 
$\theta_5$. In this case, conjunctions of body~1 and body~2 
drift at the same rate as those of body~2 and body~3
so that triple conjunction is impossible.

For the Galilean satellites, $\theta_1$ and $\theta_2$ librate
about $0^\circ$ and $180^\circ$, respectively, indicating that 
the conjunction longitude between Io and Europa is locked to Io's 
pericenter and Europa's apocenter. $\theta_3$ librates about
$0^\circ$, while $\theta_4$ circulates through $360^\circ$, which indicates
that the Europa-Ganymede conjunction is locked to Europa's
pericenter but to neither apse of Ganymede. Thus, $\theta_5$ obviously
librates about $180^\circ$, indicating a Laplace relationship.

{}

\end{document}